\documentclass[aps,nofootinbib,preprintnumbers]{revtex4}
\usepackage{CJK}
\usepackage{amsfonts}
\usepackage{amsmath}
\usepackage{amssymb}
\usepackage[english]{babel}
\usepackage{graphicx}
\usepackage{epsfig}
\usepackage{bm}
\usepackage{verbatim}
\usepackage[utf8]{inputenc}
\usepackage{booktabs}
\usepackage{multirow}
\usepackage{subfig}
\usepackage{slashed}
\usepackage{xcolor}
\usepackage[colorlinks=true,urlcolor=red,citecolor=red]{hyperref}
\usepackage[font=small]{caption}
\usepackage{float}
\usepackage{blindtext}
\usepackage{placeins}
\usepackage[utf8]{inputenc}
\usepackage{bbm}
\usepackage[colorinlistoftodos]{todonotes}

\newenvironment{Eqnarray}{\arraycolsep 0.14em\begin{eqnarray}}{\end{eqnarray}}
\newcommand{\ba}{\begin{Eqnarray}}
\newcommand{\ea}{\end{Eqnarray}}
\newcommand{\be}{\begin{equation}}
\newcommand{\ee}{\end{equation}}
\newcommand{\bal}{\begin{aligned}}
\newcommand{\eal}{\end{aligned}}
\newcommand{\bea}{\begin{eqnarray}}
\newcommand{\eea}{\end{eqnarray}}
\newcommand{\ben}{\begin{enumerate}}
\newcommand{\een}{\end{enumerate}}
\newcommand{\bit}{\begin{itemize}}
\newcommand{\eit}{\end{itemize}}
\newcommand{\bde}{\begin{widetext}}
\newcommand{\ede}{\end{widetext}}

\renewcommand{\[}{\left[}

\def\lsim{\mathrel{\rlap{\lower4pt\hbox{\hskip1pt$\sim$}}
    \raise1pt\hbox{$<$}}}
\def\gsim{\mathrel{\rlap{\lower4pt\hbox{\hskip1pt$\sim$}}
    \raise1pt\hbox{$>$}}}

\def\3211{$\mathrm{SU(3) \otimes SU(2)_L \otimes U(1)_R \otimes U(1)_{B-L}}$ }
\def\321{$\mathrm{SU(3) \otimes SU(2) \otimes U(1)}$ }
\def\422{$\mathrm{SU(4) \otimes SU(2) \otimes SU(2)_R}$ }

\newcommand{\mathsym}[1]{{}}

\definecolor{bostonuniversityred}{rgb}{0.8, 0.0, 0.0}

\input{tcilatex}
\begin{document}

\title{Left-Right model with radiative double seesaw mechanism}

\author{Paulo Areyuna C.}
\email{paulo.areyuna@sansano.usm.cl}
\affiliation{{Millennium Institute for Subatomic Physics at High-Energy Frontier
(SAPHIR), Fern\'andez Concha 700, Santiago, Chile}}
\affiliation{Departamento de Física, Facultad de Ciencias, Universidad de La Serena, Avenida Cisternas
1200, La Serena, Chile}
\author{A. E. C\'arcamo Hern\'andez}
\email{antonio.carcamo@usm.cl}
\affiliation{{Universidad T\'ecnica Federico Santa Mar\'{\i}%
a, Casilla 110-V, Valpara\'{\i}so, Chile}}
\affiliation{{Centro Cient\'{\i}fico-Tecnol\'ogico de Valpara\'{\i}so, Casilla 110-V,
Valpara\'{\i}so, Chile}}
\affiliation{{Millennium Institute for Subatomic Physics at High-Energy Frontier
(SAPHIR), Fern\'andez Concha 700, Santiago, Chile}}
\author{Vishnudath K. N.}
\email{vishnudath.neelakand@usm.cl}
\affiliation{Universidad T\'ecnica Federico Santa Mar\'{\i}a, Casilla 110-V, Valpara\'{\i}so, Chile}
\author{Sergey Kovalenko}
\email{sergey.kovalenko@unab.cl}
\affiliation{Center for Theoretical and Experimental Particle Physics - CTEPP, Facultad de Ciencias Exactas, Universidad Andres Bello, Fernandez Concha 700, Santiago, Chile}
\affiliation{{Millennium Institute for Subatomic Physics at High-Energy Frontier, SAPHIR,
Chile}}
\affiliation{{Centro Cient\'{\i}fico-Tecnol\'ogico de Valpara\'{\i}so, Casilla 110-V,
Valpara\'{\i}so, Chile}}
\author{Roman Pasechnik}
\email{roman.pasechnik@fysik.lu.se}
\affiliation{Department of Physics,\\
Lund University, SE-223 62 Lund, Sweden}
\author{Iv\'an Schmidt
\footnote{Our friend and collaborator Iv\'an Schmidt passed away during the completion of this work. He will be sorely missed.}}
\affiliation{{Departamento de F\'{\i}sica, Universidad T\'ecnica Federico Santa Mar\'{\i}%
a, Casilla 110-V, Valpara\'{\i}so, Chile}}
\affiliation{{Centro Cient\'{\i}fico-Tecnol\'ogico de Valpara\'{\i}so, Casilla 110-V,
Valpara\'{\i}so, Chile}}
\affiliation{{Millennium Institute for Subatomic Physics at High-Energy Frontier
(SAPHIR), Fern\'andez Concha 700, Santiago, Chile}}

\begin{abstract}
We propose an extended Left-Right symmetric model with an additional global symmetry $U(1)_X$, which after spontaneous symmetry breaking collapses to a residual subgroup $\mathbb{Z}_2$, ensuring that the light active neutrino masses are generated via a double seesaw mechanism at two loop level, with the Dirac submatrix arising at one loop. It also guarantees one loop level masses for the SM charged fermions lighter than the top quark and protects Dark Matter (DM) candidates of the model. To the best of our knowledge our model has the first implementation of the radiative double seesaw mechanism with the Dirac submatrix generated at one loop level. We show that the model can successfully accommodate the observed pattern of SM fermion masses as well as mixings and is compatible with the constraints arising from the muon $g-2$ anomaly, neutrinoless double beta decay and DM.
\end{abstract}

\maketitle

\section{Introduction}

The Standard Model (SM) of particle physics has been very successful as a theory of fundamental interactions, and its predictions have been confirmed to a high degree of accuracy in several experiments. Despite its great success, the SM is notoriously unable to explain a number of fundamental aspects of nature, such as the observed patterns of mass and mixing of the SM fermions, the origin of Dark Matter (DM) and the source of parity violation in electroweak (EW) interactions, to name a few.

Even though the energy scale of New Physics is unknown, current experimental searches keep constraining the popular beyond-the-SM (BSM) models pushing the scale of New Physics towards high energies. This motivates the BSM theories having experimental signatures that manifest themselves at energies much higher than the Fermi scale of EW interactions, such as for instance the Left-Right (LR) theories \cite{Pati:1974yy,Mohapatra:1974gc,Davidson:1987mh,deAlmeida:2010qb,Gu:2010xc,CarcamoHernandez:2018hst,Dekens:2014ina,Nomura:2016run,Brdar:2018sbk,Ma:2020lnm,Babu:2020bgz,Hernandez:2021uxx,Bonilla:2023wok}, which are possible low energy limits of Grand unified models such as those based on the exceptional $E_{6,8}$ groups as well as the trinification symmetry $SU(3)_{C}\times SU(3)_{L}\times SU(3)_{R}$ (see e.g.~Refs.~\cite{Babu:1985gi,Choi:2003ag,Willenbrock:2003ca,Frampton:2004vw,Kim:2004pe,Carone:2004rp,Sayre:2006ma,Leontaris:2008mm,Cauet:2010ng,Stech:2012zr,Stech:2014tla,Hetzel:2015bla,Hetzel:2015cca,Pelaggi:2015kna,Camargo-Molina:2016yqm,Reig:2016vtf,Camargo-Molina:2016bwm,Reig:2016tuk,Hati:2017aez,Camargo-Molina:2017kxd,Dong:2017zxo,Wang:2018yer,Dinh:2019jdg,Aranda:2020fkj,Morais:2020ypd,Morais:2020odg,CarcamoHernandez:2020owa,Aranda:2021bvg,Aranda:2021eyn} for previous studies of this type of models). These theories are very attractive since they include several appealing features, foremost of which is to provide a compelling explanation of parity violation in EW interactions, as a low-energy effect arising from the spontaneous breaking of the LR-symmetry at high scales. 

We are thus proposing, as a possible explanation of the aforementioned issues, an extended LR model where the top quark mass is generated at tree level whereas the remaining SM charged fermions acquire one loop level masses thanks to the virtual charged vector-like fermions and electrically neutral scalars running in the loops. In our proposed theory the tiny masses of the light active neutrinos are produced by a double seesaw mechanism at two loop level, where the Majorana and Dirac neutrino submatrices are generated at tree and one loop level, respectively. Tree level double seesaw in the context of LR model was proposed in \cite{deAlmeida:2010qb} and its phenomenological aspects have been studied in \cite{Gu:2010xc,Patra:2023ltl,Patel:2023voj}. To the best of our knowledge our model is the first implementation of the radiative double seesaw mechanism with the Dirac submatrix generated at one loop level. 

The radiative realization of the seesaw mechanisms is ensured by a conserved $\mathbb{Z}_2$ symmetry resulting from the spontaneous breaking of the global $U(1)_X$ symmetry of our model. Such preserved $\mathbb{Z}_2$ symmetry also guarantees the stability of the DM candidate, which will correspond to the lightest particle having a non trivial charge under this remnant $\mathbb{Z}_2$ symmetry. In our model, the LR gauge symmetry $SU(3)_C\times SU(2)_L\times SU(2)_R\times U(1)_{B-L}$ is augmented by the inclusion of a spontaneously broken global $U(1)_X$ group. The model fermion spectrum is enlarged with respect to the conventional LR model setup by adding charged vector like fermions and right handed Majorana neutrinos. The scalar spectrum of our theory contains one scalar bi-doublet, two $SU(2)_L$ and two $SU(2)_R$ scalar doublets as well as two electrically neutral scalar singlets. These fields are crucial for the implementation of the above mentioned radiative seesaw mechanisms that yield the observed SM charged fermion mass hierarchy and explain the tiny values of the active neutrino masses.  Furthermore, the charged vector like leptons which mediate the one loop level seesaw mechanism that yields the SM charged lepton masses and the Dirac neutrino submatrix also provide radiative contributions to the muon and electron anomalous magnetic moments. Thus, our proposed scenario provides a natural connection between the fermion mass generation mechanism and the $g-2$ muon anomaly. Besides that, our model has a rich DM phenomenology and provides testable predictions for neutrino less double beta decay.

This paper is structured as follows. In section~\ref{model} we provide a detailed explanation of the model with its particle spectrum and symmetries, which includes a comprehensive description of how the SM charged fermion mass hierarchy and the tiny active neutrino masses emerge. The consequences of the model in the quark, charged lepton and neutrino mass spectra and mixing is discussed in section~\ref{fermiomassesandmixings}. In section~\ref{gminus2muon} we analyze the implications of the model for the muon anomalous magnetic moment. Section~\ref{nunubeta} includes a detailed analysis of the implications of the model for neutrinoless double beta decay. A comprehensive analysis and discussion of the DM phenomenology is included in section~\ref{DM}. We conclude our results in section~\ref{conclusions}.

\section{Left-Right double seesaw model}
\label{model}

In what follows, we proceed by explaining the reasoning that justifies the inclusion of extra scalars, fermions and symmetries required for the implementation of the one loop seesaw mechanism that generates the masses of the SM charged fermions lighter than the top quark as well as of two loop double seesaw that produces the tiny masses of the active neutrinos with one loop induced Dirac submatrix. In our theoretical construction that is described below, the full neutrino mass matrix in the basis $\left(\nu_{L},\nu_{R}^{C},N_{R}^{C}\right)$ has the following structure: 
\begin{equation}
M_{\nu }=\left( 
\begin{array}{ccc}
0_{3\times 3} & m_{\nu D} & 0_{3\times 3} \\ 
m_{\nu D}^{T} & 0_{3\times 3} & M \\ 
0_{3\times 3} & M^{T} & M_{N}%
\end{array}%
\right) \,,  \label{Mnufull}
\end{equation}%
where $\nu_{iL}$ ($i=1,2,3$) correspond to the active neutrinos, whereas $\nu_{iR}$ and $N_{iR}$ ($i=1,2,3$) are the sterile neutrinos. Furthermore, the entries of the full neutrino mass matrix of Eq.~(\ref{Mnufull}) should satisfy the hierarchy $\left(m_{\nu D}\right)_{ij}\ll M_{ij} \ll ( M_{N})_{ij}$ ($i,j=1,2,3 $). Here, the submatrices $M$ and $M_{N}$ arise at tree level, whereas the submatrix $m_{\nu D}$ is radiatively generated at one loop level.

The theory under consideration is based on the $SU(3)_{C}\times SU(2)_{L}\times SU(2)_{R}\times U(1)_{B-L}$ gauge symmetry, which is supplemented by the inclusion of a $U(1)_{X}$ global symmetry. The latter ensures the radiative nature of the double seesaw mechanism that produces the tiny active neutrino masses, as well as the seesaw-like mechanism that generates the masses of the SM charged fermions lighter than the top quark.

In our proposed model, the top quark mass will arise from a renormalizable Yukawa operator, with an order one Yukawa coupling, i.e:
\begin{equation}
\overline{Q}_{3L}\Phi Q_{iR},\hspace{1.5cm}i=1,2,3 \,, 
\label{top}
\end{equation}
where $\Phi$ is a bidoublet scalar, whereas $Q_{3L}$ and $Q_{iR}$ ($i=1,2,3$) are $SU(2)_{L}$ and $SU(2)_{R}$ quark doublets, respectively.

The masses of the SM charged fermions lighter than the top quark can emerge
from the following operators:
\begin{eqnarray}
&&\overline{Q}_{3L}\phi_{L}B_{3R},\hspace{1.5cm}\overline{B}_{3L}\phi_{R}^{\dagger }Q_{iR},\hspace{1.5cm}\overline{B}_{3L}\eta B_{3R},\hspace{1.5cm}i,r=1,2,3 \,,  \notag \\
&&\overline{Q}_{nL}\widetilde{\phi }_{L}T_{kR},\hspace{1.5cm}\overline{T}_{kL}\widetilde{\phi }_{R}^{\dagger }Q_{iR},\hspace{1.5cm}\overline{T}_{kL}\sigma T_{kR},\hspace{1.5cm}k=1,2 \,,  \notag \\
&&\overline{Q}_{nL}\phi _{L}B_{kR},\hspace{1.5cm}\overline{B}_{kL}\phi_{R}^{\dagger }Q_{iR},\hspace{1.5cm}m_{B_{k}}\overline{B}_{kL}B_{kR}\,,  \notag
\\
&&\overline{L}_{iL}\phi_{L}E_{rR},\hspace{1.5cm}\overline{E}_{rL}\phi_{R}^{\dagger }L_{jR},\hspace{1.5cm}m_{E_{r}}\overline{E}_{rL}E_{rR} \,,
\label{SMop}
\end{eqnarray}
where $L_{iL}$ ($Q_{iL}$) and $L_{iR}$ ($Q_{iR}$) are $SU(2)_{L}$ and $SU(2)_{R}$ lepton (quark) doublets, respectively, and $\chi_{L}$, $\phi_{L}$ and $\chi_{R}$, $\phi_{R}$ are $SU(2)_{L}$ and $SU(2)_{R}$ scalar doublets, respectively. Furthermore, $T_{n}$ ($n=1,2$), $B_{i}$ and $E_{i}$ ($i=1,2,3$) are heavy vector-like up, down type quarks and charged leptons, respectively. 

It is worth mentioning that the operators of the last line of Eq.~(\ref{SMop}) give rise to a one loop level Dirac submatrix $m_{\nu D}$. Besides, the $\phi_{L}$ and $\phi_{R}$\ scalar doublets are assumed to have odd $U(1)_{X}$ charges, which will imply that they do not acquire vacuum expectation values (VEVs) as the $U(1)_{X}$ global symmetry is assumed to be spontaneously broken down to a preserved discrete symmetry. Therefore, the masses of the SM charged fermions lighter than the top quark are radiatively generated at one-loop level. The implementation of such a radiative seesaw mechanism also requires the inclusion of the heavy vector-like up-type quarks $T_{n}$ ($n=1,2$), down-type quarks $B_{i}$ and charged leptons $E_{i}$ ($i=1,2,3$) in the fermion spectrum of the model under consideration.

To generate the Majorana submatrices $M$ and $M_{N}$, the following
operators are required:
\begin{equation}
\overline{N_{iR}^{C}}\widetilde{\chi }_{R}^{\dagger }L_{jR},\hspace{1.5cm}
(M_{N})_{ij}\overline{N}_{iR}N_{jR}^{C} \,,  \label{MandMRop}
\end{equation}
where $N_{iR}$ ($i=1,2,3$) are right-handed Majorana neutrinos. Notice that the the operators of the last line of Eq.~(\ref{SMop}) and the ones given by Eq.~(\ref{MandMRop}) allow for a successful implementation of the double seesaw mechanism at two-loop level that yields the tiny active neutrino masses. This is a consequence of the fact that the light active neutrino mass matrix arising from the double seesaw mechanism will have a quadratic and linear dependence on the one loop induced Dirac $m_{\nu D}$ and Majorana $M_{N}$ submatrices, respectively.

The symmetry of the model $\mathcal{G}$ exhibits the following spontaneous breaking pattern:
\begin{eqnarray}
&&\mathcal{G}=SU(3)_{C}\times SU(2)_{L}\times SU(2)
_{R}\times U(1)_{B-L}\times U(1)_{X}  \notag \\
&&\hspace{35mm}\Downarrow v_{\sigma },v_{\eta },v_{R}  \notag \\[0.12in]
&&\hspace{15mm}SU(3)_{C}\times SU(2)_{L}\times U(1)
_{Y}\times \mathbb{Z}_{2}  \notag \\[0.12in]
&&\hspace{35mm}\Downarrow v_{1},v_{L}  \notag \\[0.12in]
&&\hspace{25mm}SU(3)_{C}\otimes U(1)_{Q}\times \mathbb{Z}_{2}
\end{eqnarray}%
We assume that the LR symmetry and the global $U(1)_{X}$ symmetry are spontaneously broken at the same energy scale, taken to be around $v_{R}\sim \mathcal{O}(10)$ TeV. 

With the scalar content specified in Table~\ref{scalars}, the global $U(1)_{X}$ symmetry is spontaneously broken down to a preserved $\mathbb{Z}_{2}$ discrete symmetry with the following parities of the model fields :
\begin{equation}
 \mathbf{Z}_2:\ \ \ \  \mathcal{P}(\Psi) =\left(-1\right) ^{X}\,,
 \end{equation}
where $X$ is the $U(1)_{X}$ charge of the field $\Psi$. This $\mathbb{Z}_2$ symmetry remains unbroken at low energies and is crucial for guaranteeing the radiative nature of the considered seesaw mechanism. The quark and leptonic spectrum of the model with their transformations under the $SU(3)_{C}\times SU(2)_{L}\times SU(2)_{R}\times U(1)_{B-L}\times U(1)_X$ group are displayed in Tables~\ref{quarks} and \ref{leptons}, respectively. 

It is worth mentioning that we have enlarged the fermion sector of the original LR model by adding only gauge singlet Majorana neutrinos $N_{R_{i}}$ ($i=1,2,3$) and vector-like exotic fermions: two exotic up-type quarks $T_{n}$($n=1,2$), three exotic down-type quarks $B_i$, three charged leptons $E_{i}$ ($i=1,2,3$). Gauge singlet and vector-like fermions do not contribute to the gauge group chiral anomaly, and hence our model is anomaly-free as the original LR model is anomaly-free by construction. Besides that, the exotic fermions of the model under consideration are assigned as singlet representations of the LR $SU(2)_{L}\times SU(2)_{R}$ gauge group.

The scalar fields of the model can be expanded as follows: 
\begin{eqnarray}
\Phi  &=&\left( 
\begin{array}{cc}
\frac{1}{\sqrt{2}}\left( v_{1}+\phi _{1R}^{0}+i\phi _{1I}^{0}\right)  & \phi
_{2}^{+} \\ 
\phi _{1}^{-} & \frac{1}{\sqrt{2}}\left( v_{2}+\phi _{2R}^{0}+i\phi
_{2I}^{0}\right) 
\end{array}%
\right) ,  \notag \\
\chi _{L} &=&\left( 
\begin{array}{c}
\chi _{L}^{+} \\ 
\frac{1}{\sqrt{2}}\left( v_{L}+\func{Re}\chi _{L}^{0}+i\func{Im}\chi
_{L}^{0}\right) 
\end{array}%
\right) ,\hspace{1cm}\chi _{R}=\left( 
\begin{array}{c}
\chi _{R}^{+} \\ 
\frac{1}{\sqrt{2}}\left( v_{R}+\func{Re}\chi _{R}^{0}+i\func{Im}\chi
_{R}^{0}\right) 
\end{array}%
\right) , \\
\phi _{L} &=&\left( 
\begin{array}{c}
\phi _{L}^{+} \\ 
\frac{1}{\sqrt{2}}\left( \func{Re}\phi _{L}^{0}+i\func{Im}\phi
_{L}^{0}\right) 
\end{array}%
\right) ,\hspace{1cm}\phi _{R}=\left( 
\begin{array}{c}
\phi _{R}^{+} \\ 
\frac{1}{\sqrt{2}}\left( \func{Re}\phi _{R}^{0}+i\func{Im}\phi
_{R}^{0}\right) 
\end{array}%
\right) ,  \notag \\
\sigma  &=&\frac{1}{\sqrt{2}}\left( v_{\sigma }+\func{Re}\sigma +i\func{Im}%
\sigma \right) ,\hspace{1cm}\eta =\frac{1}{\sqrt{2}}\left( v_{\eta }+\func{Re%
}\eta +i\func{Im}\eta \right) \,.   \notag
\end{eqnarray}
Let us note that the $SU(2)_{R}$ scalar doublet $\chi_{R}$ is required for the spontaneous breaking of the $SU(2)_{R}\times U(1)_{B-L}$ symmetry, whereas the $SU(2)_{L}$ scalar doublet $\chi _{L}$ and the $SU(2)_{L}\times SU(2)_{R}$ bi-doublet scalar $\Phi$ are needed to break the EW gauge symmetry spontaneously.

The spontaneous breaking $U(1)_X\to \mathbb{Z}_2$ is triggered by VEVs of the $\chi_R$, $\sigma$ and $\eta$ scalars. Notice that the gauge singlet scalars $\sigma$ and $\eta$ are crucial for providing tree level masses to the vector-like up-type quarks $T_{n}$ ($n=1,2$) and down-type quark $B_3$, respectively. Moreover, a successful implementation of the radiative seesaw mechanism that yields one-loop level masses for the light SM charged fermions (below the top mass) requires to include the dark $SU(2)_L$ and $SU(2)_R$ scalars $\phi_L$ and $\phi_R$, respectively. Note that the scalar fields $\phi_L$ and $\phi_R$ have odd charges under the preserved remnant of $U(1)_X$ -- the 
$\mathbb{Z}_2$ symmetry. Consequently, $\phi_L$ and $\phi_R$ do not acquire VEVs, thus, ensuring the radiative nature of the seesaw mechanism as mentioned above.
\begin{table}[tbp]
\begin{tabular}{|c|c|c|c|c|c|c|c|}
\hline
& $\Phi$  & $\chi_{L}$ & $\chi_{R}$ & $\phi_{L}$ & $\phi_{R}$ & $\sigma$  & $\eta$  \\ \hline
$SU(3)_{C}$ & $\mathbf{1}$ & $\mathbf{1}$ & $\mathbf{1}$ & $\mathbf{1}$ & 
$\mathbf{1}$ & $\mathbf{1}$ & $\mathbf{1}$ \\ \hline
$SU(2)_{L}$ & $\mathbf{2}$ & $\mathbf{2}$ & $\mathbf{1}$ & $\mathbf{2}$ & 
$\mathbf{1}$ & $\mathbf{1}$ & $\mathbf{1}$ \\ \hline
$SU(2)_{R}$ & $\mathbf{2}$ & $\mathbf{1}$ & $\mathbf{2}$ & $\mathbf{1}$ & 
$\mathbf{2}$ & $\mathbf{1}$ & $\mathbf{1}$ \\ \hline
$U(1)_{B-L}$ & $0$ & $1$ & $1$ & $1$ & $1$ & $0$ & $0$ \\ \hline
$U(1)_{X}$ & $0$ & $0$ & $-2$ & $-1$ & $1$ & $4$ & $-2$ \\ \hline
\end{tabular}
\caption{Scalar assignments under the $SU(3)_{C}\times SU(2)_{L}\times SU(2)_{R}\times U(1)_{B-L}\times U(1)_{X}$ symmetry.}
\label{scalars}
\end{table}
\begin{table}[tbp]
	\begin{tabular}{|c|c|c|c|c|c|c|c|c|c|}
		\hline
& $Q_{nL}$ & $Q_{3L}$ & $Q_{iR}$ & $T_{kL}$ & $T_{kR}$ & $B_{kL}$ & $B_{kR}$ & $B_{3L}$ & 
$B_{3R}$ \\ \hline
$SU(3)_{C}$ & $\mathbf{3}$ & $\mathbf{3}$ & $\mathbf{3}$ & $\mathbf{3}$ & 
$\mathbf{3}$ & $\mathbf{3}$ & $\mathbf{3}$ & $\mathbf{3}$ & $\mathbf{3}$ \\ \hline
$SU(2)_{L}$ & $\mathbf{2}$ & $\mathbf{2}$ & $\mathbf{1}$ & $\mathbf{1}$ & 
$\mathbf{1}$ & $\mathbf{1}$ & $\mathbf{1}$ & $\mathbf{1}$ & $\mathbf{1}$ \\ \hline
$SU(2)_{R}$ & $\mathbf{1}$ & $\mathbf{1}$ & $\mathbf{2}$ & $\mathbf{1}$ & 
$\mathbf{1}$ & $\mathbf{1}$ & $\mathbf{1}$ & $\mathbf{1}$ & $\mathbf{1}$ \\ \hline
$U(1)_{B-L}$ & $\frac{1}{3}$ & $\frac{1}{3}$ & $\frac{1}{3}$ & $\frac{4}{%
3}$ & $\frac{4}{3}$ & $-\frac{2}{3}$ & $-\frac{2}{3}$ & $-\frac{2}{3}$ & $-\frac{2}{3}$
\\ \hline
$U(1)_{X}$ & $-1$ & $1$ & $1$ & $2$ & $-2$ & $0$ & $0$ & $0$ & $2$ \\ \hline
\end{tabular}%
\caption{Quark assignments under the $SU(3)_{C}\times SU(2)_{L}\times SU(2)_{R}\times U(1)_{B-L}\times U(1)_{X}$ symmetry.}
\label{quarks}
\end{table}
\begin{table}[tbp]
\begin{tabular}{|c|c|c|c|c|c|}
\hline
& $L_{iL}$ & $L_{iR}$ & $E_{iL}$ & $E_{iR}$ & $N_{iR}$ \\ \hline
$SU(3)_{C}$ & $\mathbf{1}$ & $\mathbf{1}$ & $\mathbf{1}$ & $\mathbf{1}$ & 
$\mathbf{1}$ \\ \hline
$SU(2)_{L}$ & $\mathbf{2}$ & $\mathbf{1}$ & $\mathbf{1}$ & $\mathbf{1}$ & 
$\mathbf{1}$ \\ \hline
$SU(2)_{R}$ & $\mathbf{1}$ & $\mathbf{2}$ & $\mathbf{1}$ & $\mathbf{1}$ & 
$\mathbf{1}$ \\ \hline
$U(1)_{B-L}$ & $-1$ & $-1$ & $-2$ & $-2$ & $0$ \\ \hline
$U(1)_{X}$ & $0$ & $2$ & $1$ & $1$ & $0$\\ \hline
\end{tabular}%
\caption{Lepton assignments under the $SU(3)_{C}\times SU(2)_{L}\times SU(2)_{R}\times U(1)_{B-L}\times U(1)_{X}$ symmetry.}
\label{leptons}
\end{table}

The VEVs of the scalars $\Phi $, $\chi_{L}$ and $\chi_{R}$ read: 
\begin{equation}
\left\langle \Phi \right\rangle =\left( 
\begin{array}{cc}
\frac{v_{1}}{\sqrt{2}} & 0 \\ 
0 & \frac{v_{2}}{\sqrt{2}}%
\end{array}%
\right) ,\hspace{1.5cm}\left\langle \chi _{L}\right\rangle =\left( 
\begin{array}{c}
0 \\ 
\frac{v_{L}}{\sqrt{2}}%
\end{array}%
\right) ,\hspace{1.5cm}\left\langle \chi _{R}\right\rangle =\left( 
\begin{array}{c}
0 \\ 
\frac{v_{R}}{\sqrt{2}}%
\end{array}%
\right) \,,
\end{equation}%
where for the sake of simplicity we set $v_2=0$.

With the above particle content, the following relevant Yukawa terms arise: 
\begin{eqnarray}
-\mathcal{L}_{Y} &=&\dsum\limits_{i=1}^{3}y_{Q_{i}}\overline{Q}_{3L}\Phi
Q_{iR}+w_{B}\overline{Q}_{3L}\phi _{L}B_{3R}+\dsum\limits_{i=1}^{3}r_{i}%
\overline{B}_{3L}\phi _{R}^{\dagger }Q_{iR}+y_{B}\overline{B}_{3L}\eta B_{3R}\notag\\
&&+\dsum\limits_{n=1}^{2}\dsum\limits_{k=1}^{2}\left( x_{T}\right) _{nk}%
\overline{Q}_{nL}\widetilde{\phi }_{L}T_{kR}+\dsum\limits_{k=1}^{2}\dsum%
\limits_{i=1}^{3}\left( z_{T}\right) _{ki}\overline{T}_{kL}\widetilde{\phi }%
_{R}^{\dagger }Q_{iR}+\dsum\limits_{k=1}^{2}y_{T_{k}}\overline{T}_{kL}\sigma
T_{kR} \notag\\
&&+\dsum\limits_{n=1}^{2}\dsum\limits_{k=1}^{2}\left( x_{B}\right) _{nk}%
\overline{Q}_{nL}\phi
_{L}B_{kR}+\dsum\limits_{k=1}^{2}\dsum\limits_{i=1}^{3}\left( z_{B}\right)
_{ki}\overline{B}_{kL}\phi _{R}^{\dagger
}Q_{iR}+\dsum\limits_{k=1}^{2}m_{B_{k}}\overline{B}_{kL}B_{kR} \notag\\
&&+\dsum\limits_{i=1}^{3}\dsum\limits_{r=1}^{3}\left( x_{E}\right) _{ir}%
\overline{L}_{iL}\phi
_{L}E_{rR}+\dsum\limits_{r=1}^{3}\dsum\limits_{j=1}^{3}\left( z_{E}\right)
_{rj}\overline{E}_{rL}\phi _{R}^{\dagger
}L_{jR}+\dsum\limits_{r=1}^{3}m_{E_{r}}\overline{E}_{rL}E_{rR} \notag\\
&&+\dsum\limits_{i=1}^{3}\dsum\limits_{j=1}^{3}\left( x_{N}\right) _{ij}%
\overline{N_{iR}^{C}}\widetilde{\chi }_{R}^{\dagger
}L_{jR}+\dsum\limits_{i=1}^{3}\dsum\limits_{j=1}^{3}\left( M_{N}\right) _{ij}%
\overline{N}_{iR}N_{jR}^{C}+H.c.  
\label{Ly}
\end{eqnarray}
\begin{figure}[]
\includegraphics[width=8cm, height=5cm]{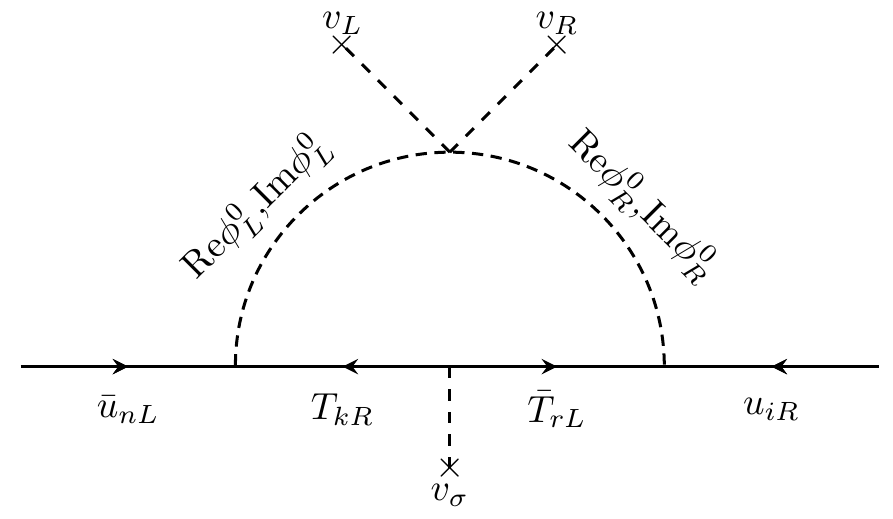} %
\includegraphics[width=8cm, height=4cm]{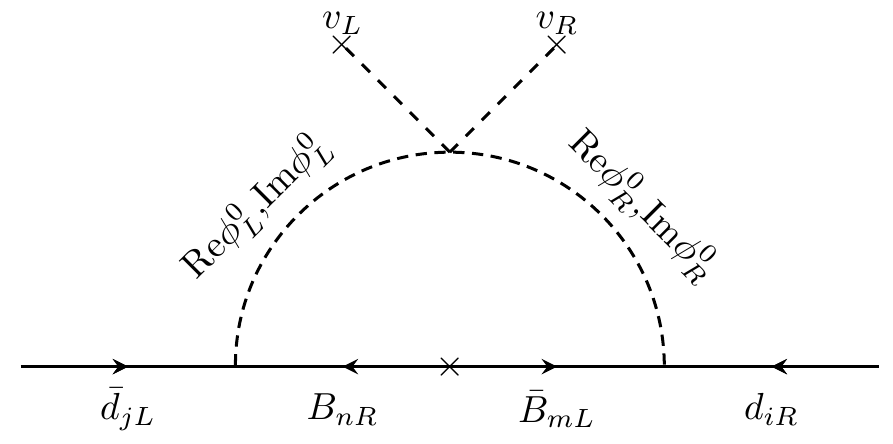}\newline
\caption{Feynman diagrams contributing to the entries of the quark mass matrices. Here, $k,r=1,2$ and $i,j,n,m=1,2,3$.}
\label{Diagramsquarks}
\end{figure}
\begin{figure}[]
\includegraphics[width=8cm, height=4cm]{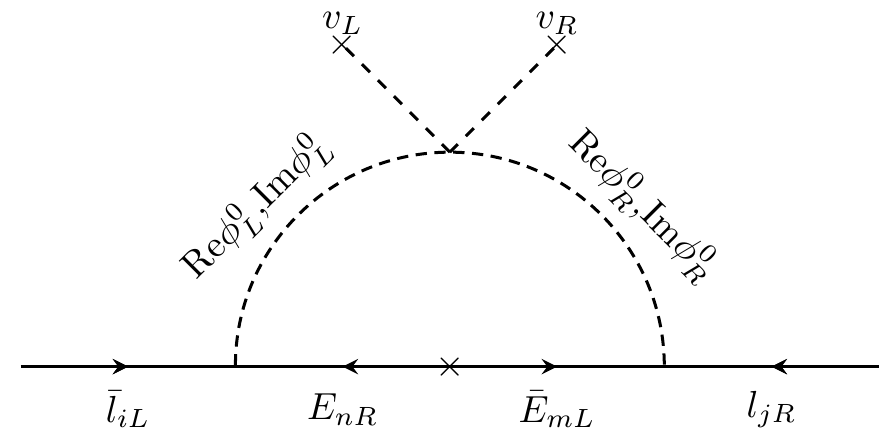} %
\includegraphics[width=8cm, height=4cm]{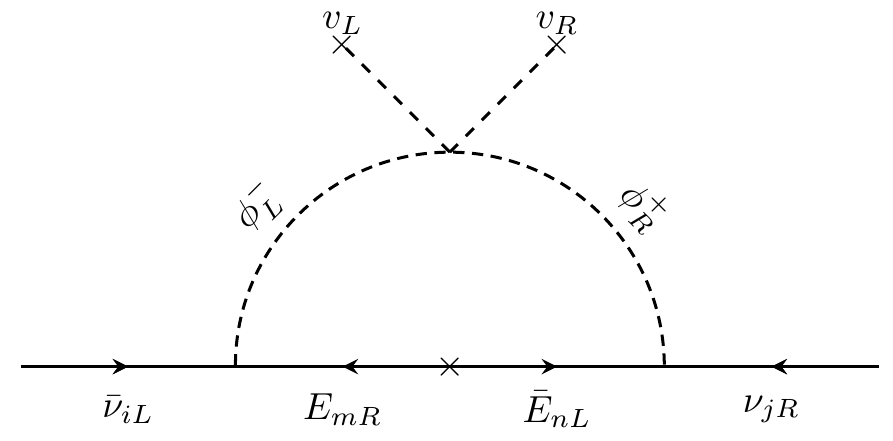}\newline
\caption{Feynman diagrams contributing to the entries of the charged lepton mass matrices and Dirac neutrino submatrix $m_{\nu D}$. Here, $i,j,n,m=1,2,3$.}
\label{Diagramsleptons}
\end{figure}

\section{Fermion masses and mixing}
\label{fermiomassesandmixings}

In this section, we discuss the implications of our model for the SM fermion masses and mixing. Since the global $U(1)_X$ symmetry is spontaneously broken down to the preserved $\mathbb{Z}_2$ discrete symmetry, the light SM charged fermions (except for the top quark) acquire one loop level masses, whereas the top quark and exotic fermions obtain their masses at tree level. The tiny masses of the active neutrinos are generated via a double seesaw mechanism at two loop level, with the Dirac neutrino submatrix being radiatively generated at one loop. The Feynman diagrams contributing to the entries of the quark mass matrices are shown in Figure~\ref{Diagramsquarks}. The entries of the charged lepton mass matrices and Dirac neutrino submatrix $m_{\nu D}$ arise from the Feynman diagrams displayed in Figure~\ref{Diagramsleptons}. From the Yukawa interactions (\ref{Ly}), we find that the mass matrices for SM charged fermions are given by:
\begin{eqnarray}
\left( M_{U}\right) _{ni} &=&\dsum\limits_{k=1}^{2}\frac{\left( x_{T}\right)
_{nk}\left( z_{T}\right) _{ki}m_{T_{k}}}{16\pi ^{2}}\left\{ \left[ f\left(
m_{H_{1}}^{2},m_{T_{k}}^{2}\right) -f\left(
m_{H_{2}}^{2},m_{T_{k}}^{2}\right) \right] \sin 2\theta _{H}+\left[ f\left(
m_{A_{1}}^{2},m_{T_{k}}^{2}\right) -f\left(
m_{A_{2}}^{2},m_{T_{k}}^{2}\right) \right] \sin 2\theta _{A}\right\} , 
\notag \\
\left( M_{U}\right) _{3i} &=&y_{Q_{i}}\frac{v_{1}}{\sqrt{2}},\hspace{1.5cm}n=1,2,\hspace{1.5cm}i=1,2,3 \,.
\label{MU}
\end{eqnarray}
\begin{eqnarray}
\left( M_{D}\right) _{ni} &=&\dsum\limits_{k=1}^{2}\frac{\left( x_{B}\right)
_{nk}\left( z_{B}\right) _{ki}m_{B_{k}}}{16\pi ^{2}}\left\{ \left[ f\left(
m_{H_{1}}^{2},m_{B_{k}}^{2}\right) -f\left(
m_{H_{2}}^{2},m_{B_{k}}^{2}\right) \right] \sin 2\theta _{H}+\left[ f\left(
m_{A_{1}}^{2},m_{B_{k}}^{2}\right) -f\left(
m_{A_{2}}^{2},m_{B_{k}}^{2}\right) \right] \sin 2\theta _{A}\right\} \,, 
\notag \\
\left( M_{D}\right) _{3i} &=&\frac{w_{B}r_{i}m_{B_{3}}}{16\pi ^{2}}\left\{ %
\left[ f\left( m_{H_{1}}^{2},m_{B_{3}}^{2}\right) -f\left(
m_{H_{2}}^{2},m_{B_{3}}^{2}\right) \right] \sin 2\theta _{H}+\left[ f\left(
m_{A_{1}}^{2},m_{B_{3}}^{2}\right) -f\left(
m_{A_{2}}^{2},m_{B_{3}}^{2}\right) \right] \sin 2\theta _{A}\right\} \,,
\label{MD}
\end{eqnarray}
\begin{equation}
\left( M_{l}\right) _{ij}=\dsum\limits_{r=1}^{3}\frac{\left( x_{E}\right)
_{ir}\left( z_{E}\right) _{rj}m_{E_{r}}}{16\pi ^{2}}\left\{ \left[ f\left(
m_{H_{1}}^{2},m_{E_{r}}^{2}\right) -f\left(
m_{H_{2}}^{2},m_{E_{r}}^{2}\right) \right] \sin 2\theta _{H}+\left[ f\left(
m_{A_{1}}^{2},m_{E_{r}}^{2}\right) -f\left(
m_{A_{2}}^{2},m_{E_{r}}^{2}\right) \right] \sin 2\theta _{A}\right\} \,,
\label{Ml}
\end{equation}
with $f\left(m_{1}^{2},m_{2}^{2}\right) $ being a loop function defined as, \begin{equation}
F\left(m_{1}^{2},m_{2}^{2}\right) =\frac{m_{1}^{2}}{m_{1}^{2}-m_{2}^{2}}\ln \left( 
\frac{m_{1}^{2}}{m_{2}^{2}}\right) \,.
\end{equation}%
Furthermore, $H_1$, $H_2$ and $A_1$, $A_2$ are the physical CP-even and CP-odd dark scalars, respectively. They are given by: 
\begin{eqnarray}
&&\left( 
\begin{array}{c}
H_{1} \\ 
H_{2}%
\end{array}%
\right)=R_H\left( 
\begin{array}{c}
\func{Re}\phi _{L}^{0} \\ 
\func{Re}\phi _{R}^{0}%
\end{array}%
\right)=\left( 
\begin{array}{cc}
\cos \theta _{H} & \sin \theta _{H} \\ 
-\sin \theta _{H} & \cos \theta _{H}%
\end{array}%
\right) \left( 
\begin{array}{c}
\func{Re}\phi _{L}^{0} \\ 
\func{Re}\phi _{R}^{0}%
\end{array}%
\right) \, , \notag\\
&&\left( 
\begin{array}{c}
\ A_{1} \\ 
A_{2}%
\end{array}%
\right)=R_A\left( 
\begin{array}{c}
\func{Im}\phi _{L}^{0} \\ 
\func{Im}\phi _{R}^{0}%
\end{array}%
\right)=\left( 
\begin{array}{cc}
\cos \theta _{A} & \sin \theta _{A} \\ 
-\sin \theta _{A} & \cos \theta _{A}%
\end{array}%
\right) \left( 
\begin{array}{c}
\func{Im}\phi _{L}^{0} \\ 
\func{Im}\phi _{R}^{0}%
\end{array}%
\right) \,,
\end{eqnarray}
whereas $H_{1}^{\pm }$ and $H_{2}^{\pm }$ are the electrically charged scalars arising from the linear combinations of the electrically charged components of the dark $SU(2)_L$ and $SU(2)_R$ scalar doublets $\phi_L$ and $\phi_R$. These fields are defined as follows:
\begin{equation}
\left( 
\begin{array}{c}
H_{1}^{\pm } \\ 
H_{2}^{\pm }%
\end{array}%
\right) =\left( 
\begin{array}{cc}
\cos \theta  & \sin \theta  \\ 
-\sin \theta  & \cos \theta 
\end{array}%
\right) \left( 
\begin{array}{c}
\func{Re}\phi _{L}^{\pm } \\ 
\func{Re}\phi _{R}^{\pm }%
\end{array}%
\right) \,. 
\end{equation}
For $\theta \ll \theta_{H},\theta_{A}$, the SM down-type quarks and
charged lepton mass matrices can be parameterized as follows: 
\begin{eqnarray}
\widetilde{M}_{D} &\simeq &\widetilde{A}_{D}J_{D}^{-1}\widetilde{B}_{D}^{T},%
\hspace{0.7cm}\hspace{0.7cm}J_{D}=\left( 
\begin{array}{ccc}
\frac{1}{16\pi ^{2}}m_{B_{1}}K_{D}^{(1)} & 0 & 0 \\ 
0 & \frac{1}{16\pi ^{2}}m_{B_{2}}K_{D}^{(2)} & 0 \\ 
0 & 0 & \frac{1}{16\pi ^{2}}m_{B_{3}}K_{D}^{(3)}%
\end{array}%
\right) \,,  \label{Md} \\
\widetilde{M}_{E} &=&\widetilde{A}_{E}J_{E}^{-1}\widetilde{B}_{E}^{T},%
\hspace{0.7cm}\hspace{0.7cm}J_{E}=\left( 
\begin{array}{ccc}
\frac{1}{16\pi ^{2}}m_{E_{1}}K_{E}^{(1)} & 0 & 0 \\ 
0 & \frac{1}{16\pi ^{2}}m_{E_{2}}K_{E}^{(2)} & 0 \\ 
0 & 0 & \frac{1}{16\pi ^{2}}m_{E_{3}}K_{E}^{(3)}%
\end{array}%
\right) \,,  \label{Ml}
\end{eqnarray}%
where
\begin{eqnarray}
K_{D}^{\left( i\right) } &=&\left\{ \left[ f\left(
m_{H_{1}}^{2},m_{B_{i}}^{2}\right) -f\left(
m_{H_{2}}^{2},m_{B_{i}}^{2}\right) \right] \sin 2\theta _{H}+\left[ f\left(
m_{A_{1}}^{2},m_{B_{i}}^{2}\right) -f\left(
m_{A_{2}}^{2},m_{B_{i}}^{2}\right) \right] \sin 2\theta _{A}\right\} \, , \\
K_{E}^{\left( i\right) } &=&\left\{ \left[ f\left(
m_{H_{1}}^{2},m_{E_{i}}^{2}\right) -f\left(
m_{H_{2}}^{2},m_{E_{i}}^{2}\right) \right] \sin 2\theta _{H}+\left[ f\left(
m_{A_{1}}^{2},m_{E_{i}}^{2}\right) -f\left(
m_{A_{2}}^{2},m_{E_{i}}^{2}\right) \right] \sin 2\theta _{A}\right\} \,,
\end{eqnarray}%
\begin{eqnarray}
\widetilde{A}_{D} &=&V_{L}^{\left( D\right) }M_{d}^{\frac{1}{2}}J_{D}^{\frac{%
1}{2}},\hspace{0.7cm}\hspace{0.7cm}\widetilde{B}_{D}=V_{R}^{\left( D\right)
}M_{d}^{\frac{1}{2}}J_{D}^{\frac{1}{2}},\hspace{0.7cm}\hspace{0.7cm}%
M_{d}=\left( 
\begin{array}{ccc}
m_{d} & 0 & 0 \\ 
0 & m_{s} & 0 \\ 
0 & 0 & m_{b}%
\end{array}%
\right) \, , \\
\widetilde{A}_{E} &=&V_{L}^{\left( E\right) }M_{l}^{\frac{1}{2}}J_{E}^{\frac{%
1}{2}},\hspace{0.7cm}\hspace{0.7cm}\widetilde{B}_{E}=V_{R}^{\left( E\right)
}M_{l}^{\frac{1}{2}}J_{E}^{\frac{1}{2}},\hspace{0.7cm}\hspace{0.7cm}%
M_{l}=\left( 
\begin{array}{ccc}
m_{e} & 0 & 0 \\ 
0 & m_{\mu } & 0 \\ 
0 & 0 & m_{\tau }%
\end{array}%
\right) \,.
\end{eqnarray}

In what follows, we show that our model can successfully reproduce
the following experimental values of the quark masses \cite{Xing:2020ijf},
the Cabibbo–Kobayashi–Maskawa (CKM) parameters \cite{ParticleDataGroup:2022pth} and the charged lepton
masses \cite{ParticleDataGroup:2022pth}: 
\begin{eqnarray}
&&m_{u}(MeV)=1.24\pm 0.22,\hspace{3mm}m_{d}(MeV)=2.69\pm 0.19 \,,\hspace{3mm}%
m_{s}(MeV)=53.5\pm 4.6,  \notag \\
&&m_{c}(GeV)=0.63\pm 0.02,\hspace{3mm}m_{t}(GeV)=172.9\pm 0.4 \,,\hspace{3mm}%
m_{b}(GeV)=2.86\pm 0.03,\hspace{3mm}  \notag \\
&&\sin \theta _{12}=0.2245\pm 0.00044,\hspace{3mm}\sin \theta
_{23}=0.0421\pm 0.00076,\hspace{3mm}\sin \theta _{13}=0.00365\pm 0.00012 \,, 
\notag \\
&&J=\left( 3.18\pm 0.15\right) \times 10^{-5}\,,
\label{eq:Qsector-observables} \\
&&m_{e}(MeV)=0.4883266\pm 0.0000017,\ \ \ \ m_{\mu }(MeV)=102.87267\pm
0.00021,\ \ \ \ m_{\tau }(MeV)=1747.43\pm 0.12 \,,  \notag
\end{eqnarray}%
where $J$ is the Jarlskog parameter. By solving the eigenvalue problem for
the SM charged fermion mass matrices, we find a sample solution for the parameters
that reproduces the values in Eq.~(\ref{eq:Qsector-observables}). It is
given by: 
\begin{eqnarray}
\theta _{H} &\simeq &236^{\circ },\hspace{0.8cm}\theta _{A}\simeq 293^{\circ
},\hspace{0.8cm}m_{H_{1}}\simeq 2.1\mbox{TeV},\hspace{0.8cm}m_{H_{2}}\simeq 2%
\mbox{TeV},\hspace{0.8cm}m_{A_{1}}\simeq 2\mbox{TeV},\hspace{0.8cm}%
m_{A_{2}}\simeq 1.9\mbox{TeV}\notag \\
y_{Q_{1}} &\simeq &0.135,\hspace{0.9cm}y_{Q_{2}}\simeq 0.104,\hspace{0.9cm}%
y_{Q_{3}}\simeq 1.0,\hspace{0.9cm}w_{B}\simeq 2.236,\hspace{0.9cm}v_{1}=240%
\mbox{GeV},\hspace{0.9cm}k=1,2,\hspace{0.9cm}i=1,2,3  \notag \\
r_{1} &\simeq &-0.002-0.011i,\hspace{0.9cm}r_{2}\simeq 0.137,\hspace{0.9cm}%
r_{3}\simeq 3.242,\hspace{0.8cm}m_{T_{k}}=m_{B_{i}}=2\mbox{TeV},\hspace{0.8cm}m_{E_{1}}\simeq 0.1\mbox{TeV} \,,\notag \\
x_{T} &=&\left( 
\begin{array}{cc}
-0.313437-0.0963115i & 0.0926613\,+0.0490709i \\ 
0.459331\,+0.0454953i & 0.539547\,-0.0495055i \\ 
& 
\end{array}%
\right) \, ,  \notag \\
z_{T} &=&\left(
\begin{array}{ccc}
 0.0642028\, +0.0211999 i & 0.752348\, +0.168139 i & 0.052363\, +0.025907 i \\
 0.260579\, +0.000447435 i & 2.50643\, +0.0233949 i & 0.271241\, -0.00158314 i \\
\end{array}
\right) \,,  \notag \\
x_{B} &=&\left( 
\begin{array}{cc}
-0.39389-0.0334656i & -0.0764586+0.011812i \\ 
0.271125\,+0.0211376i & 0.706797\,-0.0163946i \\ 
& 
\end{array}%
\right) \, , \notag \\
z_{B} &=&\left(
\begin{array}{ccc}
 -0.0286419+0.00196572 i & 0.0349194\, -0.000666478 i & -0.0881341-0.0517311 i \\
 -0.0298275-0.00219832 i & 0.19754\, +0.0037934 i & 0.483457\, +0.0336031 i \\
\end{array}
\right) \,,\notag \\
x_{E} &=&\left(
\begin{array}{ccc}
 0.127945\, +0.0977866 i & 0.0203055\, +0.00240605 i & -1.84886-2.06063 i \\
 0.10833\, +0.0284267 i & -0.538754-0.291119 i & 3.24659\, +1.52185 i \\
 0.0351325\, -0.00885166 i & 1.82844\, +0.0488151 i & 1.22257\, -0.0154056 i \\
\end{array}
\right),\hspace{0.8cm}m_{E_{2}}\simeq 0.3\mbox{TeV} \,, \notag \\
z_{E} &=&\left(
\begin{array}{ccc}
 -0.0508021-0.0239322 i & -0.00516279+0.00609679 i & 0.186023\, -0.0443448 i \\
 -1.60123-0.429952 i & -0.844886-0.074173 i & -0.472732+0.171525 i \\
 0.569156\, -1.91922 i & -0.365163+4.19418 i & -0.337404-0.412488 i \\
\end{array}
\right),\hspace{0.8cm}m_{E_{3}}\simeq 1\mbox{TeV} \,.
\end{eqnarray}

The indicated values effectively accommodate the measured SM charged fermion masses and CKM parameters. The charged fermion Yukawa couplings exhibit a moderate hierarchy, with their magnitudes falling within the range $\left[ 10^{-2},1\right]$. Besides that, the electrically charged fermionic seesaw messengers have masses at the TeV scale. Despite the moderate hierarchy observed in the Yukawa couplings, this scenario is notably superior to that of the SM, where a hierarchy of approximately five orders of magnitude is evident within the charged fermion sector. As to the charged lepton sector, in our numerical analysis we consider charged exotic lepton masses in the sub TeV-TeV range, which enables us to accurately accommodate the experimental values of the muon and electron anomalous magnetic moments. Moreover, it is noteworthy that the effective Yukawa couplings are directly linked to the product of two other dimensionless couplings, so a moderate hierarchy in those couplings can lead to a quadratically larger hierarchy in the effective couplings, which in turn facilitates the explanation of the observed pattern of SM charged fermion masses and quark mixing angles. Small quark mixing angles are ascribed to the hierarchical nature of the rows of the resulting SM quark mass matrices in our model. It is worth mentioning that all rows of the SM quark mass matrices, excepting the third row of the SM up-type quark mass matrix, are radiatively generated at one loop level, thus implying that they are directly proportional to the product of two dimensionless Yukawa couplings, as indicated by Eqs.~(\ref{MU}) and (\ref{MD}).

Regarding the neutrino sector, we find that the neutrino Yukawa interactions give rise to the following neutrino mass terms: 
\begin{equation}
-\mathcal{L}_{mass}^{\left( \nu \right) }=\frac{1}{2}\left( 
\begin{array}{ccc}
\overline{\nu _{L}^{C}} & \overline{\nu _{R}} & \overline{N_{R}}%
\end{array}%
\right) M_{\nu }\left( 
\begin{array}{c}
\nu _{L} \\ 
\nu _{R}^{C} \\ 
N_{R}^{C}%
\end{array}%
\right) + H.c. \,,  \label{Lnu}
\end{equation}%
where the neutrino mass matrix reads: 
\begin{equation}
M_{\nu }=\left( 
\begin{array}{ccc}
0_{3\times 3} & m_{\nu D} & 0_{3\times 3} \\ 
m_{\nu D}^{T} & 0_{3\times 3} & M \\ 
0_{3\times 3} & M^{T} & M_{N}%
\end{array}%
\right) \, ,  \label{Mnu}
\end{equation}
with the entries of the Majorana $M$ and Dirac $m_{\nu D}$ submatrices are generated at tree and one loop levels, respectively. The entries of the Majorana submatrix $M$ arise from the first term of the last line of Eq.~(\ref{Ly}) whereas the entries of the Dirac $m_{\nu D}$ submatrix are found by computing the Feynman diagram of the right panel of Figure~\ref{Diagramsleptons}. These matrix entries are given by 
\begin{eqnarray}
\left( m_{\nu D}\right) _{ij} &=&\dsum\limits_{r=1}^{3}\frac{\left(
x_{E}\right) _{ir}\left( z_{E}\right) _{rj}m_{E_{r}}}{16\pi ^{2}}\left[
f\left( m_{H_{1}^{\pm }}^{2},m_{E_{r}}^{2}\right) -f\left( m_{H_{2}^{\pm
}}^{2},m_{E_{r}}^{2}\right) \right] \sin 2\theta \,,\hspace{0.7cm}\hspace{0.7cm%
}  \notag \\
M_{ij} &=&\left( x_{N}\right) _{ij}\frac{v_{R}}{\sqrt{2}} \,,\hspace{0.7cm}%
\hspace{0.7cm}i,j,r=1,2,3\hspace{0.7cm}\hspace{0.7cm} \,.
\end{eqnarray}

The active light neutrino masses are generated from a double seesaw mechanism \cite{Barr:2003nn}, and the physical neutrino mass matrices are given by:
\begin{eqnarray}
M_{\nu }^{(1)} &=&m_{\nu D}\left( M^{T}\right)
^{-1}M_{N}M^{-1}m_{\nu D}^{T} \,,  \label{Mnu1} \\
M_{\nu }^{(2)} &=&-MM_{N}^{-1}M^{T} \,,\hspace{1cm}\hspace{1cm}
\label{Mnu2} \\
M_{\nu }^{(3)} &=&M_{N} \,,  \label{Mnu3}
\end{eqnarray}
To make sure that we are in the correct parameter space region  that reproduces the observed light neutrino masses and mixing, we use the modified Casas-Ibarra parametrization~\cite{Casas:2001sr,Ibarra:2003up} for $m_{\nu D}$, 
\begin{equation}
m_{\nu D} = U_{\rm PMNS}(M_\nu^{(1)})_{\rm diag}^{1/2} R M_N^{-1/2} M^T \,. \label{CI}
\end{equation}
In the above equation, $U_{\rm PMNS}$ is the $3\times 3$ ontecorvo–Maki–Nakagawa–Sakata (PMNS) mixing matrix~\cite{ParticleDataGroup:2022pth} and $R$ is a general complex orthogonal matrix. The Yukawa couplings and masses entering Eq.~\ref{Mnu1} can be chosen to reproduce the values of $m_{\nu D}$ as obtained from Eq.~\ref{CI}. For simplicity, we take the matrices $M_N$ and $M$ proportional to the identity matrix, i.e.~$M_N = m_N I_{3\times 3}$ and $M = m I_{3\times 3}$.

\section{$g-2$ muon anomaly}
\label{gminus2muon}

\begin{figure}[tbp]
\centering
\includegraphics[width=8.5cm, height=5cm]{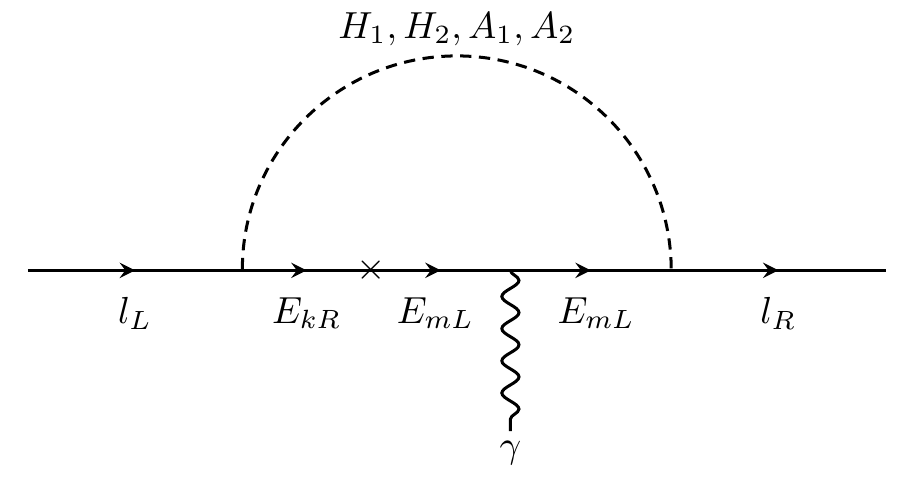}\includegraphics[width=8.5cm, height=5cm]{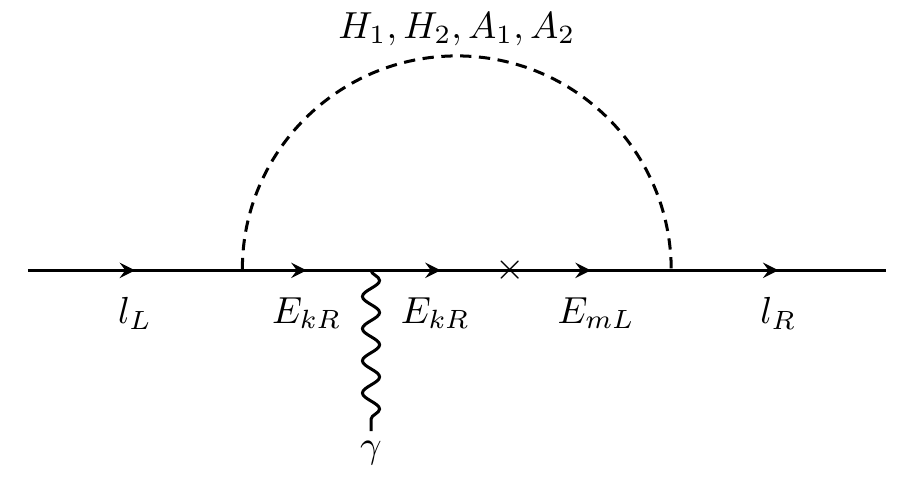}
\caption{The Feynman diagrams contributing to the anomalous magnetic moment of the muon.}
\label{gminus2Feyndiagram}
\end{figure}

In this section, we will analyze the implications of our model in the muon anomalous magnetic moment. The contributions to the muon anomalous magnetic moment in our model are shown in Fig.~\ref{gminus2Feyndiagram}. They lead to the following deviation from the SM value for this observable:
\begin{eqnarray}
\Delta a_{\mu } &=&\dsum\limits_{k=1}^{3}\frac{\func{Re}\left( \beta
_{2k}\gamma _{k2}^{\ast }\right) m_{\mu }^{2}}{8\pi ^{2}}\left[ \left(
R_{H}\right) _{11}\left( R_{H}\right) _{12}I_{S}^{\left( \mu \right) }\left(
m_{E_{k}},m_{H_{1}}\right) +\left( R_{H}\right) _{21}\left( R_{H}\right)
_{22}I_{S}^{\left( \mu \right) }\left( m_{E_{k}},m_{H_{2}}\right) \right] 
\notag \\
&&+\dsum\limits_{k=1}^{3}\frac{\func{Re}\left( \beta _{2k}\gamma _{k2}^{\ast
}\right) m_{\mu }^{2}}{8\pi ^{2}}\left[ \left( R_{A}\right) _{11}\left(
R_{A}\right) _{12}I_{P}^{\left( \mu \right) }\left(
m_{E_{k}},m_{A_{1}}\right) +\left( R_{A}\right) _{21}\left( R_{A}\right)
_{22}I_{P}^{\left( \mu \right) }\left( m_{E_{k}},m_{A_{2}}\right) \right] \,.
\notag
\end{eqnarray}
Here, the loop function $I_{S\left( P\right) }^{\left(\mu \right)
}\left( m_{E},m\right) $ has the form \cite{Diaz:2002uk,Jegerlehner:2009ry,Kelso:2014qka,Lindner:2016bgg,Kowalska:2017iqv}: 
\begin{equation}
I_{S\left( P\right) }^{\left( e,\mu \right) }\left( m_{E},m_{S}\right)
=\int_{0}^{1}\frac{x^{2}\left( 1-x\pm \frac{m_{E}}{m_{\mu }}\right) }{%
m_{\mu }^{2}x^{2}+\left( m_{E}^{2}-m_{\mu }^{2}\right) x+m_{S,P}^{2}\left(
1-x\right) }dx \,,
\end{equation}
and the dimensionless parameters $\beta _{1k}$, $\beta _{2k}$, $\gamma _{k1}$, $\gamma _{k2}$, $\kappa _{1}$, $\kappa _{2}$, $\vartheta _{1}$, $\vartheta_{2}$ are given by: 
\begin{eqnarray}
\beta _{1k} &=&\dsum\limits_{i=1}^{3}\left( x_{E}\right) _{ik}\left(
V_{lL}^{\dagger }\right) _{1i},\hspace{0.7cm}\hspace{0.7cm}\gamma
_{k1}=\dsum\limits_{j=1}^{3}\left( z_{E}\right) _{kj}\left( V_{lR}\right)
_{j1} \,, \\
\beta _{2k} &=&\dsum\limits_{i=1}^{3}\left( x_{E}\right) _{ik}\left(
V_{lL}^{\dagger }\right) _{2i},\hspace{0.7cm}\hspace{0.7cm}\gamma
_{k2}=\dsum\limits_{j=1}^{3}\left( z_{E}\right) _{kj}\left( V_{lR}\right)
_{j2} \,,
\end{eqnarray}
where $V_{lL}$ and $V_{lR}$ are the rotation matrices that diagonalize $M_{l}$ according to the relation: 
\begin{equation}
V_{lL}^{\dagger }M_{l}V_{lR}=\mathrm{diag}\left( m_{e},m_{\mu },m_{\tau
}\right)
\end{equation}
\begin{figure}[tbp]
\centering
\includegraphics[width=14cm, height=10cm]{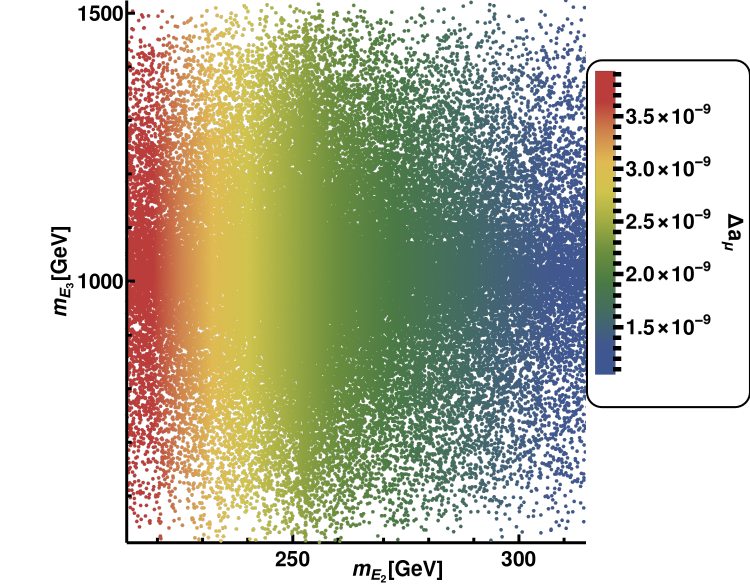}
\caption{Allowed parameter space in the $m_{E_2}-m_{E_3}$ consistent with the experimental constraints on the muon $g-2$ anomaly.}
\label{gminus2}
\end{figure}

Considering that the muon anomalous magnetic moment is constrained to be in the range \cite{Muong-2:2023cdq}, 
\begin{eqnarray}
\left( \Delta a_{\mu }\right) _{\exp } &=&\left( 2.49\pm 0.48\right) \times 10^{-9}
\end{eqnarray}
we plot in Figure~\ref{gminus2} the allowed parameter space in the $m_{E_2}-m_{E_3}$ consistent with the experimental constraints on muon $g-2$ anomaly. In our numerical analysis we have used the extended Casas-Ibarra parametrization \cite{Casas:2001sr,Ibarra:2003up} for the SM charged lepton mass matrix given in Eq.~(\ref{Ml}), to guarantee that the obtained points of the model parameter space consistent with the muon $g-2$ are also in excellent agreement with the experimental values of the SM charged lepton masses. Indeed, as shown in Figure~\ref{gminus2}, our model successfully accommodates the experimental values of the muon $g-2$ in consistency with the measured values of the SM charged fermion masses and mixing parameters.

\section{Neutrinoless double beta decay}
\label{nunubeta}

Another important feature of the LR model is that there can be a number of New Physics contributions to the neutrinoless double beta decay ($0\nu\beta\beta$), coming from right-handed currents and LR mixing, particularly, when the New Physics particles are at the TeV scale \cite{Hirsch:1996qw,Tello:2010am,Chakrabortty:2012mh,Barry:2013xxa,
Dev:2014xea,Awasthi:2015ota,
Bambhaniya:2015ipg,Bonilla:2016fqd,Awasthi:2016kbk,Deppisch:2017vne,
Borah:2017ldt,Borgohain:2017akh,Goswami:2020loc,Patra:2023ltl}. These contributions correspond to the short-range mechanism, whereas the one due to the standard light neutrino exchange correspond to the long range mechanism. These extra contributions can affect the predictions for the effective Majorana mass, which can be directly probed in $0\nu\beta\beta$ experiments. Assuming that there are no large light-heavy neutrino as well as LR gauge boson mixings, the mean half lifetime for $0\nu\beta\beta$ is given as~\cite{Tello:2010am},
\be 
\frac{1}{T_{1/2}^{0\nu}} = G_{01}^{0\nu} \Big( |\mathcal M_\nu^{0\nu} \eta_\nu|^2 +  |\mathcal M_N^{0\nu} \eta_R|^2 \Big)\,,
\ee
where, $\mathcal M_\nu^{0\nu}$ and $\mathcal M_N^{0\nu}$ are the nuclear matrix elements for the light- and heavy-neutrino exchanges, respectively, $G_{01}^{0\nu}$ is the phase space factor, and $\eta_\nu$ and $\eta_R$ represent the left-handed and right-handed amplitudes, respectively, and are given as,
\be 
\eta_\nu = \frac{1}{m_e}\sum_i U_{ei}^2 m_i \,\,\,\,;\,\,\,\, 
\eta_R = m_p \Big( \frac{M_{W_L}}{M_{W_R}} \Big)^4 \sum_i \frac{V^2_{ei} }{M_{\nu_i}^{(2)}} \,.
\ee 
Here, $m_e$, $m_p$ and $M_{W_R}$ are the masses of electron, proton and the $SU(2)_R$ gauge boson, respectively, $M_{\nu_i}^{(2)}$ is the heavy neutrino mass whereas the matrix $V$ corresponds to the mixing between active light neutrinos and heavy neutrinos. Thus, the expression for the effective Majorana neutrino mass can be written as,
\be  
m_{\beta\beta} = \sqrt{\Big|\sum_i U_{ei}^2 m_i \Big|^2 + \Big|\langle p^2 \rangle \Big( \frac{M_{W_L}}{M_{W_R}} \Big)^4 \sum_i \frac{V^2_{ei} }{M_{\nu_i}^{(2)}} \Big|^2 } \,.
\label{mee}
\ee
Here, $\,\, \langle p^2 \rangle =  |m_e m_p \mathcal M_N^{0\nu}/\mathcal M_\nu^{0\nu}|$ encapsulates the contribution due to the nuclear matrix element.
\begin{figure}[tbh]
\begin{tabular}{cc}
\includegraphics[width=0.5\textwidth]{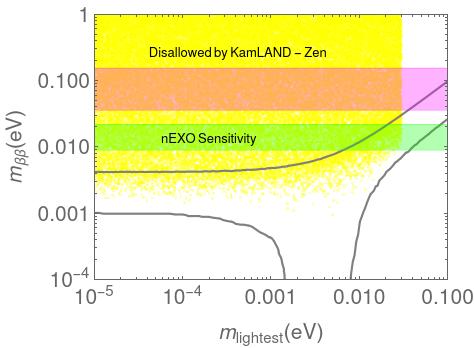}
\includegraphics[width=0.5\textwidth]{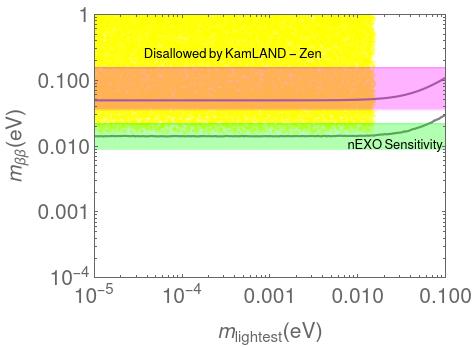}
\end{tabular}
\caption{Predictions for $m_{\beta\beta}$ in our model are shown by the yellow points. The left and the right panels are for the normal hierarchy (NH) and the inverted hierarchy (IH), respectively. The regions within the solid gray lines correspond to the predictions in the SM with three light Majorana neutrinos. The region above the magenta band is disfavored by the constraints from KamLAND-Zen~\cite{Agostini:2018tnm}. The horizontal green band corresponds to the future sensitivity of nEXO~\cite{Kharusi:2018eqi}.}
\label{fig0nbb1}
\end{figure}

In Fig.~\ref{fig0nbb1}, we have shown the predictions for $m_{\beta\beta}$ in  our model by the yellow points. The left panel corresponds to the normal hierarchy (NH) of the active light neutrinos whereas the right panel corresponds to the inverted hierarchy (IH). The regions within the solid gray lines correspond to the predictions in the SM with three light Majorana neutrinos. In generating these plots, we have used the Casas-Ibarra parametrization given by Eq.~(\ref{Ml}) for the Dirac mass matrix $m_{\nu D}$. The light neutrino mixing angles, mass-squared differences and Dirac CP phase are varied in the $3\sigma$ ranges whereas the Majorana phases are varied in the range $0-\pi$. The mass matrices $M$ and $M_N$ are taken to be diagonal and proportional to the identity matrix with their entries varying in the ranges $10-10^4$ GeV and $10^4-10^{12}$ GeV, respectively. $M_{W_R}$ is fixed to be 5 TeV. The region above the magenta band is disfavored by the constraints arising from KamLAND-Zen~\cite{Agostini:2018tnm} and the horizontal green band corresponds to the future sensitivity of nEXO~\cite{Kharusi:2018eqi}. The width of these bands are due to the uncertainty in the nuclear matrix elements. From this figure, we can see that the cancellation region of $m_{\beta \beta}$ that is present in the NH case of SM $+$ $3$ light Majorana neutrinos disappears in the LR model, for the parameter region that we have considered. Also, most of the parameter space in the IH case will be probed by nEXO in the future.

\begin{figure}[tbh]
\begin{tabular}{cc}
\includegraphics[width=0.5\textwidth]{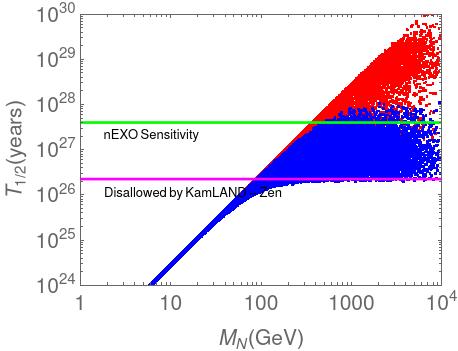}
\includegraphics[width=0.5\textwidth]{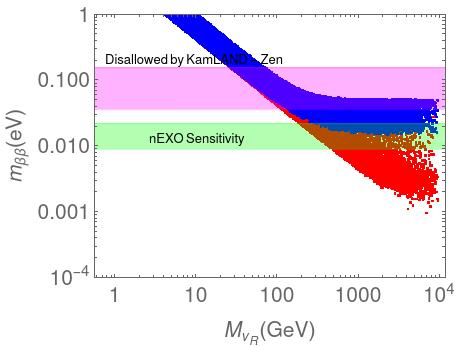}
\end{tabular}
\caption{Predictions for $T_{1/2}$ of $0\nu\beta\beta$ (left) and $m_{\beta\beta}$ (right) in  our model as a function of the heavy neutrino mass. The red and blue points correspond to NH and IH, respectively.}
\label{fig0nbb2}
\end{figure}

In Fig.~\ref{fig0nbb2}, we have shown the predictions for $T_{1/2}$ of $0\nu\beta\beta$ (left) and $m_{\beta\beta}$ (right) in  our model as a function of the heavy neutrino mass. The red and blue points correspond to NH and IH, respectively. The bound from KamLAND-Zen indicates that the heavy neutrino mass has to be $\gsim$ 80 GeV. A non-observation of $0\nu\beta\beta$ by KamLAND-Zen in the future can further raise this bound to $\sim 350$ GeV.

\section{Dark Matter phenomenology}
\label{DM}

As mentioned before, the model consists of a residual $\mathbb{Z}_2$ symmetry arising from the $U(1)_X$ symmetry breaking, under which the scalars $\phi_L$ and $\phi_R$ acquire an odd charge. This residual symmetry opens the possibility for solving the DM problem in the framework of this model. In order to study the cosmological implications of the model, we need to consider a simplified scenario. Assuming a large scale for the $U(1)_X$ symmetry breaking, we can neglect the singlet scalars. Additionally, we can integrate out the exotic fermion sector, obtaining SM-like Yukawa interactions between SM fermions and the extra scalars.

The scalar potential of the $\mathbb{Z}_2$-even sector (i.e.~$\chi_L,\chi_R,\Phi$) has been studied in Ref.~\cite{Ashry:2013loa}. Therefore, we considered the \texttt{Feynrules} \cite{fr1,fr2} files generated by the authors of this reference and extended the scalar potential by adding the $\mathbb{Z}_2$-odd doublets:
\begin{equation}\label{ldark}
\begin{split}
    \mathcal{L}_{\rm dark}=&(D_\mu \phi_L)^\dagger D^\mu \phi_L+(D_\mu \phi_R)^\dagger D^\mu \phi_R -\mu_L^2\phi_L^\dagger\phi_L-\mu_R^2\phi_R^\dagger\phi_R-\kappa(\phi_R^\dagger \phi_R)(\phi_L^\dagger \phi_L)\\
&
    -\lambda_{L2}(\phi_L^\dagger \phi_L)(\chi_L^\dagger \chi_L)-\lambda_{L3}(\chi_L^\dagger \phi_L)(\phi_L^\dagger \chi_L)-\lambda_{L4}[(\phi_L^\dagger \chi_L)(\phi_L^\dagger \chi_L)+\text{h.c.}]\\
&
      -\lambda_{R2}(\phi_R^\dagger \phi_R)(\chi_R^\dagger \chi_R)-\lambda_{R3}(\chi_R^\dagger \phi_R)(\phi_R^\dagger \chi_R)-\lambda_{R4}[(\phi_R^\dagger \chi_R)(\phi_R^\dagger \chi_R)+\text{h.c.}]\\
&  
   -\lambda_{32}(\phi_R^\dagger \phi_R)(\chi_L^\dagger \chi_L) -\lambda_{33}(\phi_L^\dagger \phi_L)(\chi_R^\dagger \chi_R) \,.
    \end{split}
\end{equation}
It is worth mentioning that Ref.~\cite{Ashry:2013loa} imposes $v_1=0$ in the bi-doublet, while the previous calculations in our work considered $v_2=0$. Nevertheless, this choice does not affect DM phenomenology (it changes the labeling scheme for the $\mathbb{Z}_2$-even particles, but does not change DM interactions). For this section, we will consider $v_2=0$. 

In this section, we neglected the terms of the scalar potential that generate mixing between $\mathbb{Z}_2$-odd particles. As will become evident in the next subsection, the choice of ignoring these mixings does not have an impact on the results and makes it easier to transfer the dependence from the couplings to the physical masses. Under this setup, the mass spectrum of the scalars coming from $\phi_L$ has the following form:
\begin{equation}
    \begin{split}
       M_{Re\phi_L^0}^2&=\mu_L^2+\frac{\lambda_{L5}v_R^2}{2}+\frac{v_L^2}{2}(\lambda_{L2}+\lambda_{L3}+\lambda_{L4})\\
       M_{Im\phi_L^0}^2&=\mu_L^2+\frac{\lambda_{L5}v_R^2}{2}+\frac{v_L^2}{2}(\lambda_{L2}+\lambda_{L3}-\lambda_{L4})\\
       M_{\phi_L^+}^2&=\mu_L^2+\frac{\lambda_{L5}v_R^2}{2}+\frac{v_L^2}{2}\lambda_{L2} \,.
    \end{split}
\end{equation}
The mass spectrum for $\phi_R$ has an analogous form, which is obtained by simply changing the subscript. In order to facilitate parameter scans, it is convenient to treat the masses as independent parameters and redefine the couplings in the following way:
\begin{equation}
\begin{split}
    \mu_L^2&= M_{Re\phi_L^0}^2-\frac{v_L^2}{2}\lambda_{L}-\frac{\lambda_{L5}v_R^2}{2}\\
    \lambda_{L2}&=\lambda_{L}+2\frac{M_{\phi_L^+}^2-M_{Re\phi_L^0}^2}{v_L^2}\\
    \lambda_{L3}&=\frac{M_{Re\phi_L^0}^2+M_{Im\phi_L^0}^2-2M_{\phi_L^+}^2}{v_L^2}\\ 
    \lambda_{L4}&=\frac{M_{Re\phi_L^0}^2-M_{Im\phi_L^0}^2}{v_L^2} \,.
\end{split}
\end{equation}
where $\lambda_{L}=\lambda_{L2}+\lambda_{L3}+\lambda_{L4}$ and $\lambda_{R}=\lambda_{R2}+\lambda_{R3}+\lambda_{R4}$. As can be seen, the couplings depend on the mass splitting between the $\phi_L$ components. For the sake of simplicity, we are focused on the highly degenerate scenario, where $M_{Re\phi_R^0}=M_{Im\phi_R^0}=M_{\phi_R^+}=M_{Im\phi_L^0}=M_{\phi_L^+}=M_{Re\phi_L^0}+\delta$, making the real part of $\phi_L^0$ the DM candidate. Additionally, we set $\lambda_{L}=\lambda_R=\lambda_{R5}=0.1$ as control parameters. In this scenario, DM phenomenology is determined by $M_{Re\phi_L^0},\delta$, and $\lambda_{L5}$.

\subsection{Relic density}

We used micrOMEGAs \cite{micromegas1,micromegas2,micromegas3} to compute the relic density assuming a freeze-out scenario, as well as to calculate the direct and indirect detection rates. Due to the uncertainties arising from our theoretical error, we consider a loose criterion for the definition of saturation. For this work, we consider a saturation of DM relic density if $0.11\leq \Omega h^2 \leq 0.12$.

Relic density as a function of $M_{Re\phi_{L}^0}$ is shown in Figure~\ref{relic}. It is worth to note that the rich scalar structure of the model generates violent drops due to the contribution of additional scalars to the annihilation cross section, avoiding overabundance at low masses. On the other hand, there is a suppression proportional to the mass splitting related to the couplings between dark doublets and the $\mathbb{Z}_2$-even scalars. While the asymptotic behavior of the relic density is highly dependent on the mass splitting, saturation is achieved in a highly degenerate scenario. This degeneracy implies that all particles in the dark sector contribute to $\langle\sigma\cdot v\rangle$, making the effect of mixings inside the dark sector negligible.
\begin{figure}
    \centering
\includegraphics[width=0.6\textwidth]{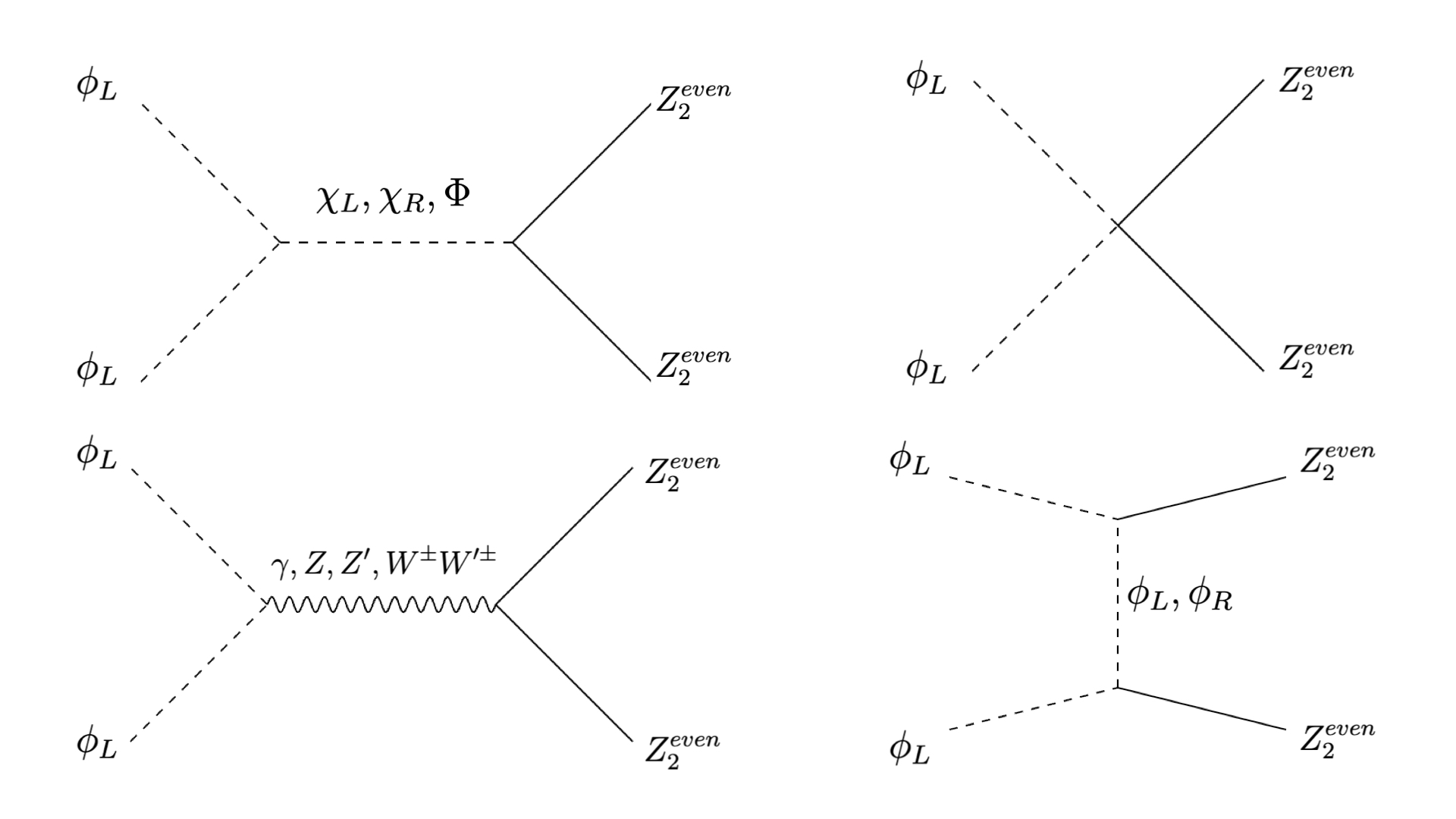}
    \caption{Main annihilation channels for DM in the early universe. Here, $\mathbb{Z}_2^{\rm even}$ stands for all particles that are even under the $\mathbb{Z}_2$ symmetry, including all particles in the SM as well as the new scalars.}
    \label{diagrams}
\end{figure}
\begin{figure}[!h]
    \includegraphics[width=0.45\textwidth]{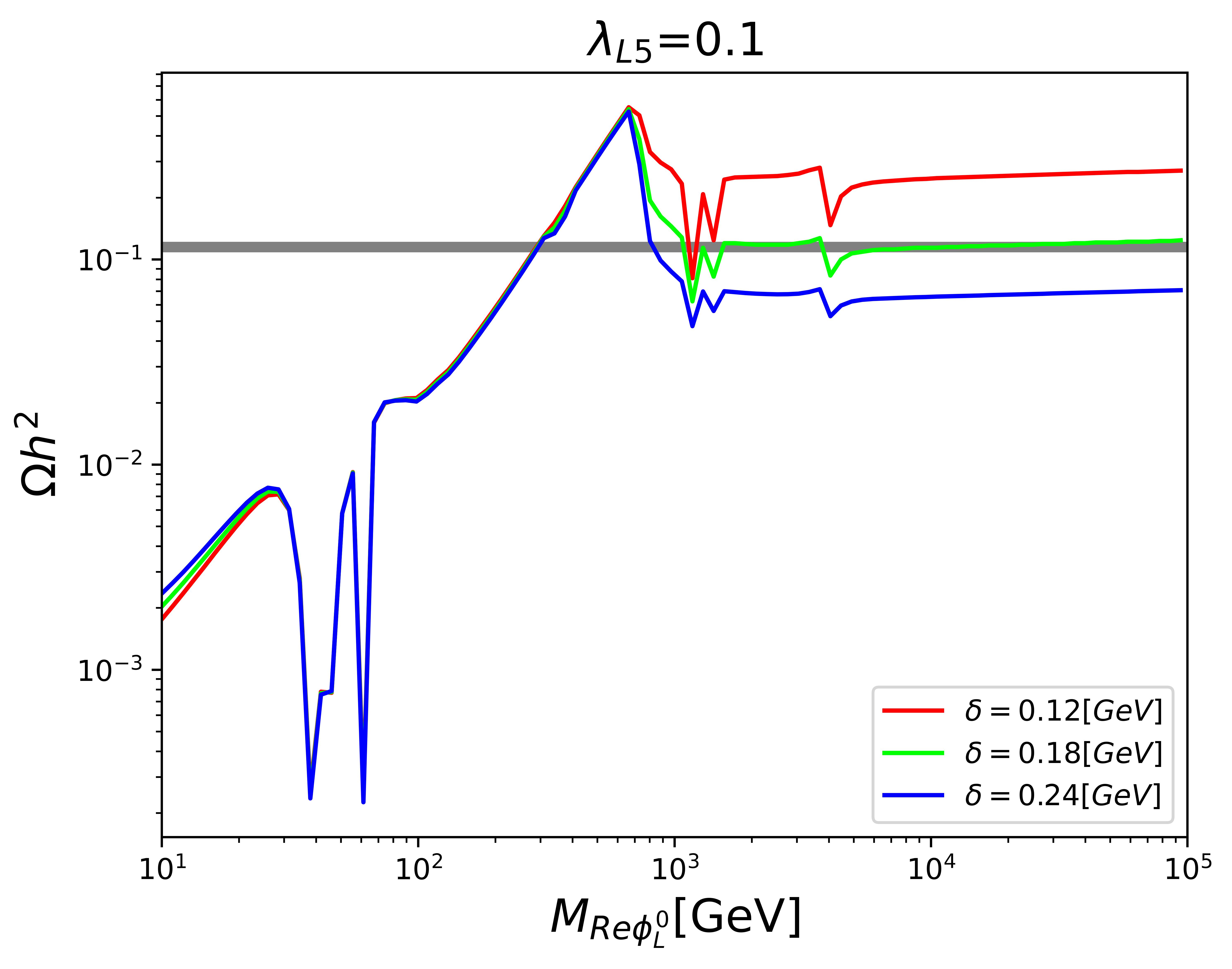}
    \includegraphics[width=0.45\textwidth]{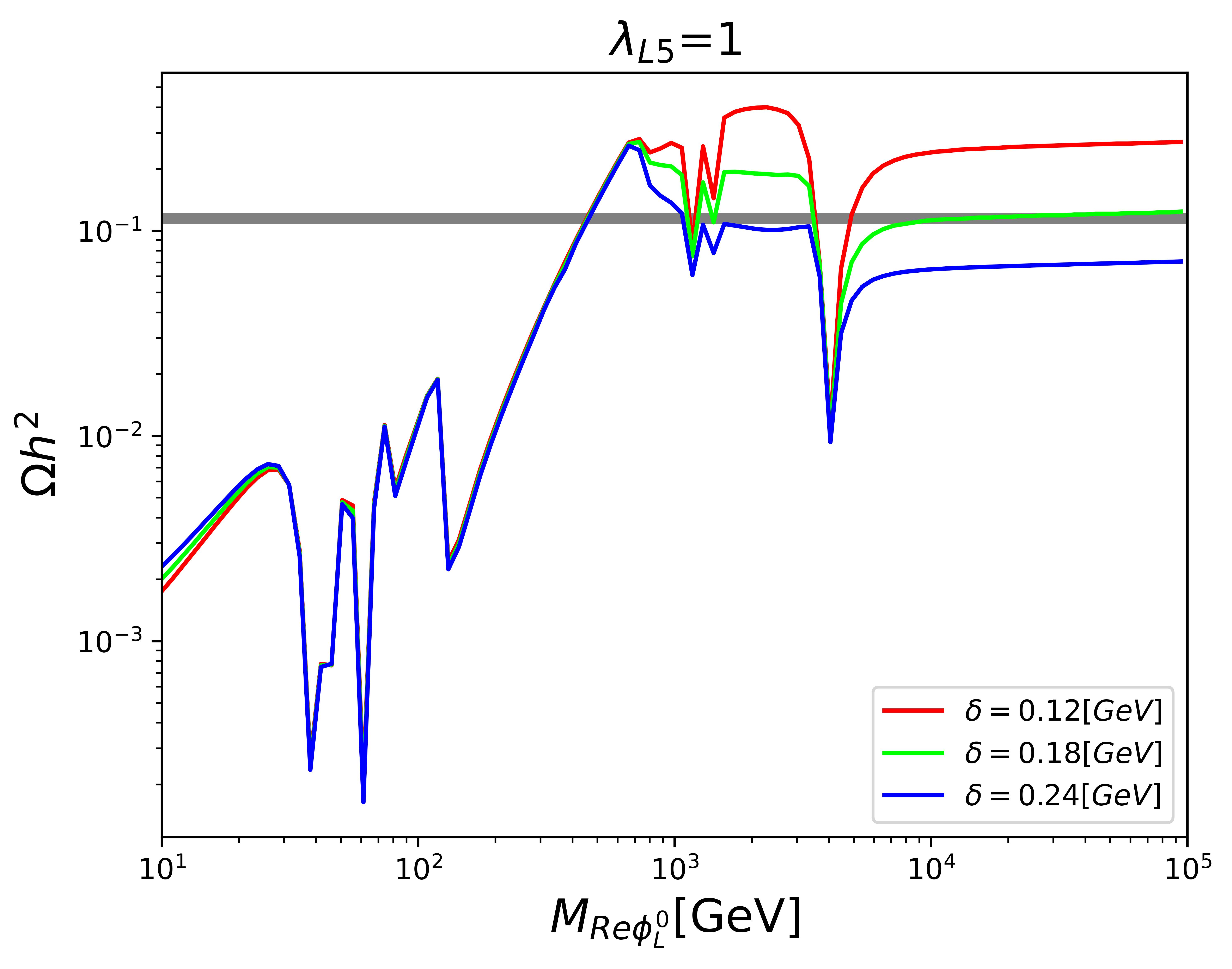}
\caption{Relic density as a function of the DM mass for different values of the scalar mass splitting $\delta$. The left panel is for $\lambda_{L5} = 0.1$ whereas the right panel is for $\lambda_{L5} = 1$.}
\label{relic}
\end{figure}
\begin{figure}[!h]
    \includegraphics[width=0.45\textwidth]{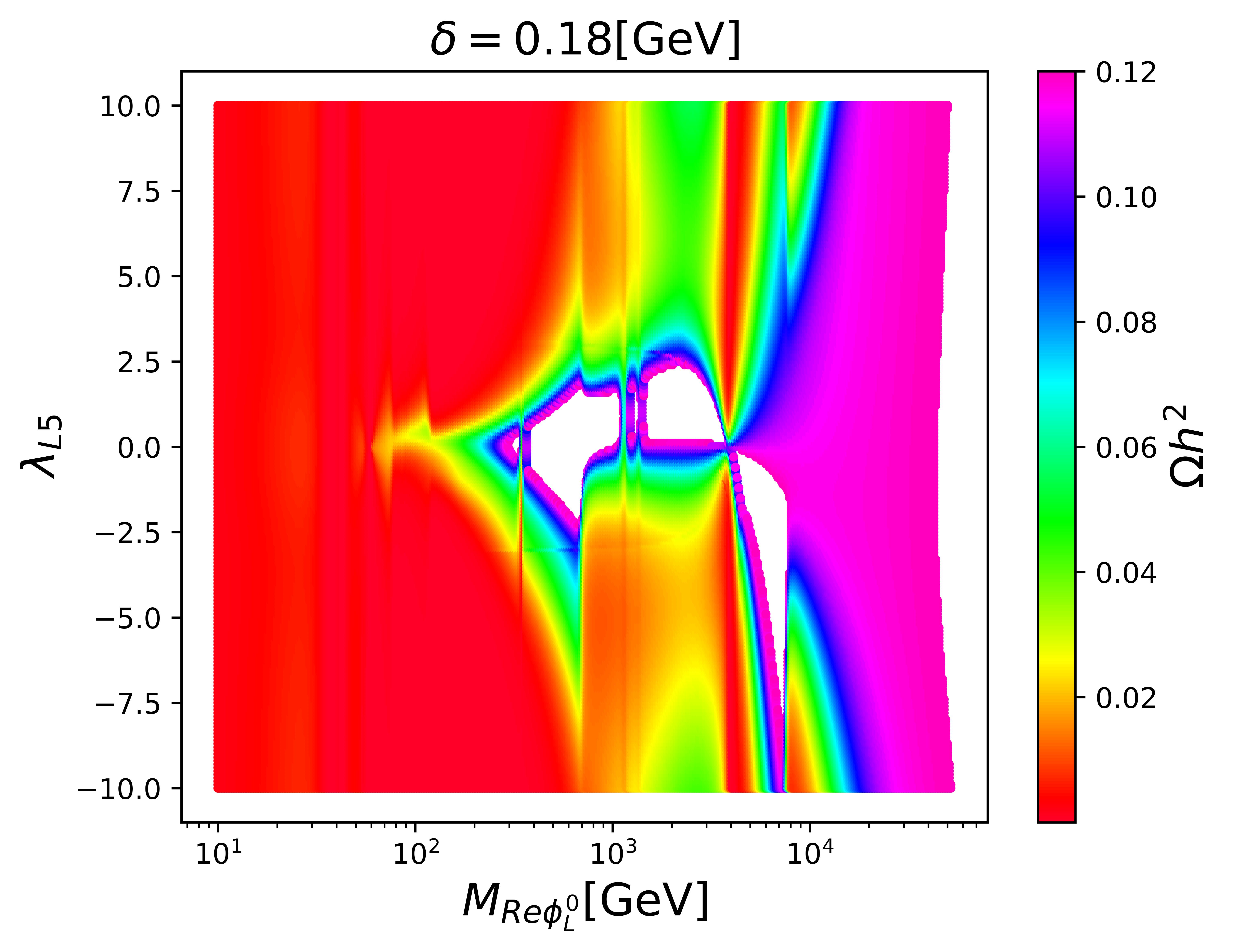}
    \includegraphics[width=0.45\textwidth]{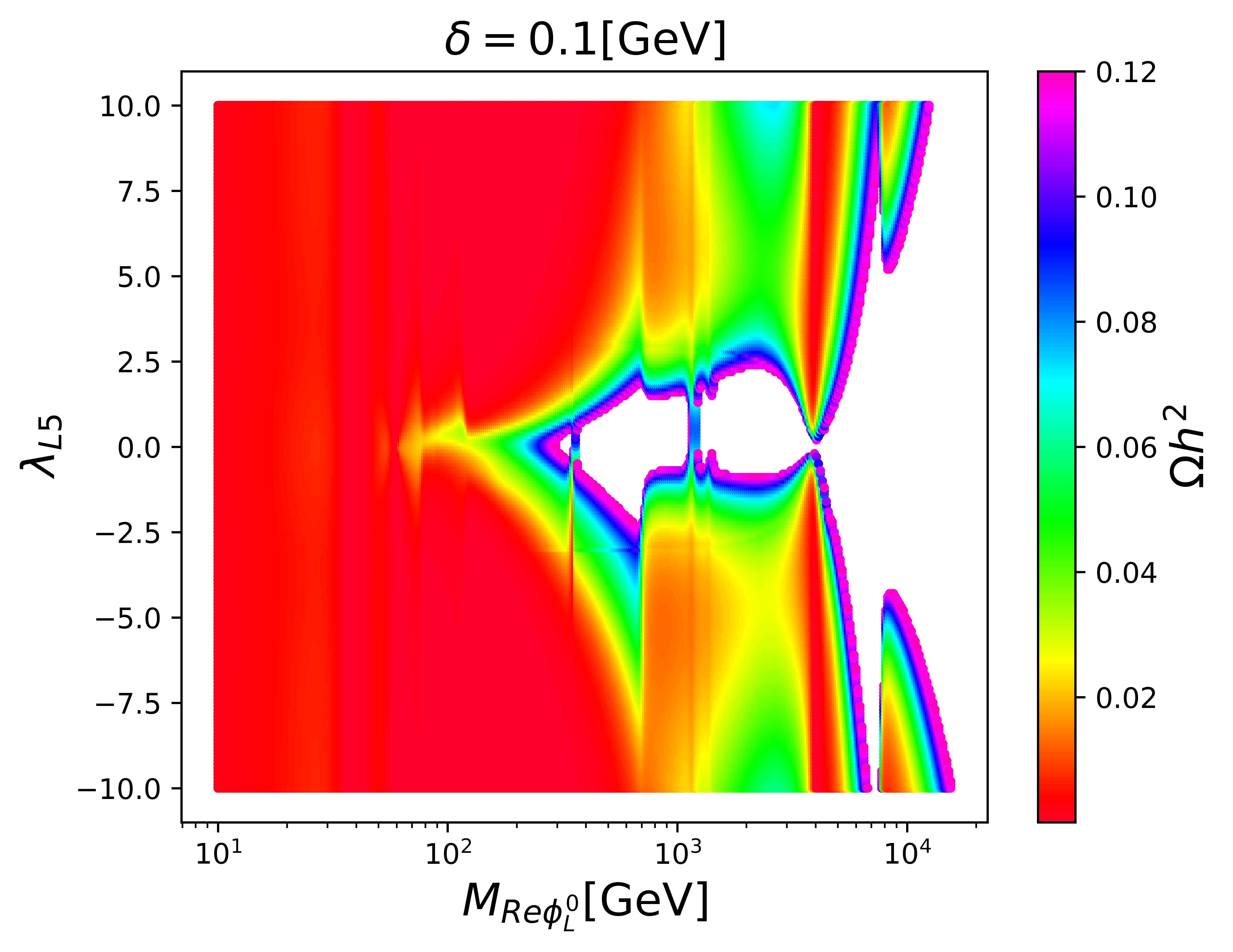}
\caption{Parameter scan for relic density in the $M_{Re\phi_L^0}-\lambda_{L5}$ plane. The left and right panels are for $\delta = 0.18$ and $0.1$ GeV, respectively. White regions represent points where the DM is over abundant.}
\label{relic_scan}
\end{figure}
\begin{figure}
    \centering
    \includegraphics[width=0.5\textwidth]{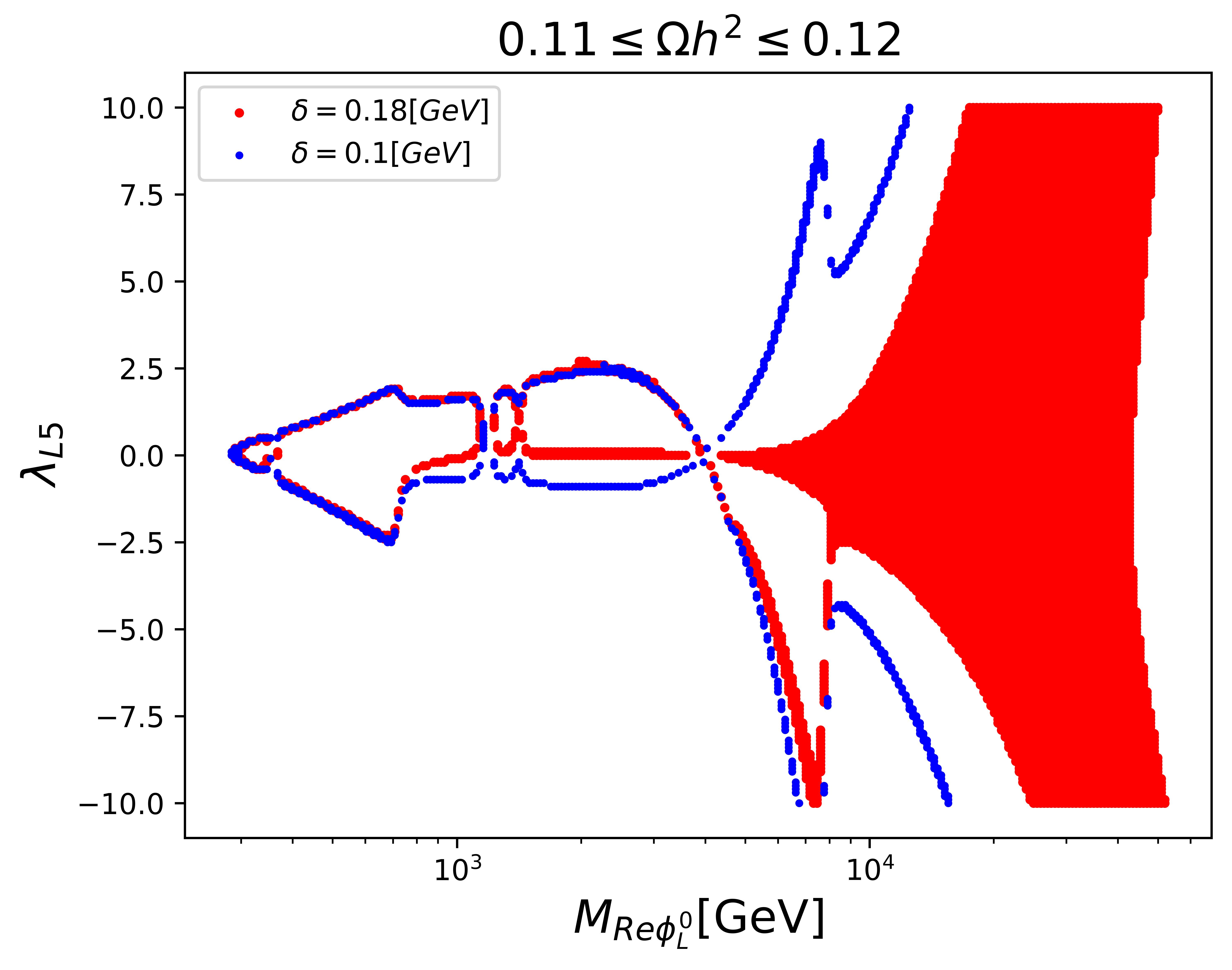}
    \caption{Parameter space points that saturate the DM relic density.}
    \label{fig:enter-label}
\end{figure}

\subsection{Indirect detection}

\begin{figure}[!h]
    \centering
    \includegraphics[width=0.45\textwidth]{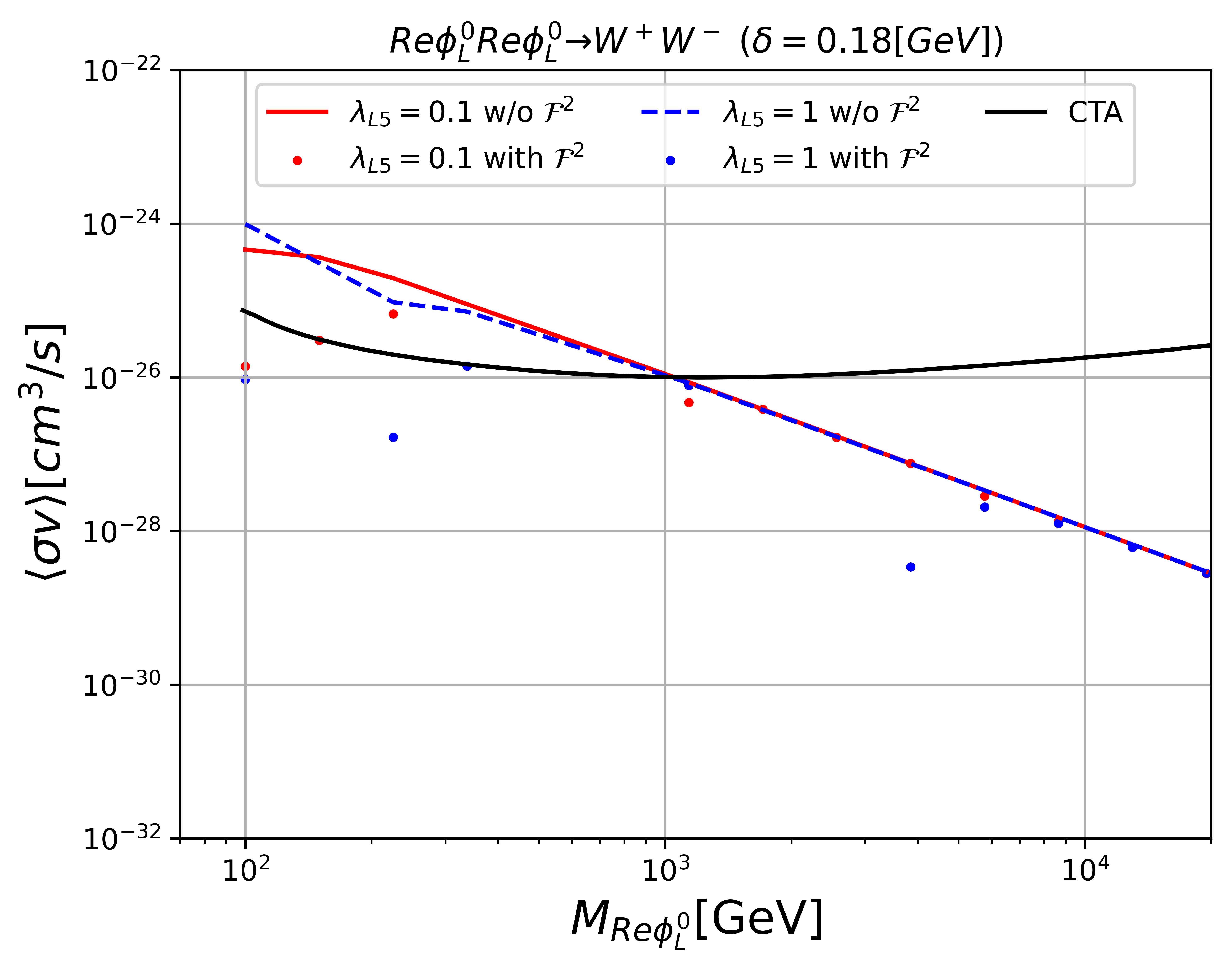}
    \caption{Annihilation cross section compared with the expected sensitivity at CTA.}
    \label{indirect}
\end{figure}

Annihilation of two DM particles in the center of the galaxy could be observed with CTA \cite{ctalims}. This collaboration has determined the expected sensitivity to the DM annihilation process as a function of the DM mass. We computed the thermally averaged annihilation cross section for $Re \phi_L^0 Re \phi_L^0 \to W^+ W^-$. However, the CTA sensitivity is obtained by assuming that the DM relic abundance is achieved by only one type of particle. Since our results are generally under-abundant, we defined the following scale factor:
\begin{equation}
    \mathcal{F}=\frac{\Omega h^2}{0.12} \,.
\end{equation}
With this scale factor, the effective annihilation cross section can be expressed as
\begin{equation}
    \langle \sigma \cdot v\rangle_{\rm eff} =\mathcal{F}^2 \langle \sigma \cdot v\rangle \,.
\end{equation}
We disregard parameter space points with $\Omega h^2>0.12$. The results can be seen in Figure~\ref{indirect}. As can be seen, a small part of parameter space could be probed at CTA, for DM mass of the order of hundreds of GeV. On the other hand, DM annihilation into the $\mathbb{Z}_{2}$-even scalars could be possible, but they would decay rapidly making a more diffuse signal. That type of processes require more sophisticated calculations that are beyond of the scope of this work and should be studied elsewhere.

\subsection{Direct detection}

\begin{figure}
    \centering
    \includegraphics[width=0.45\textwidth]{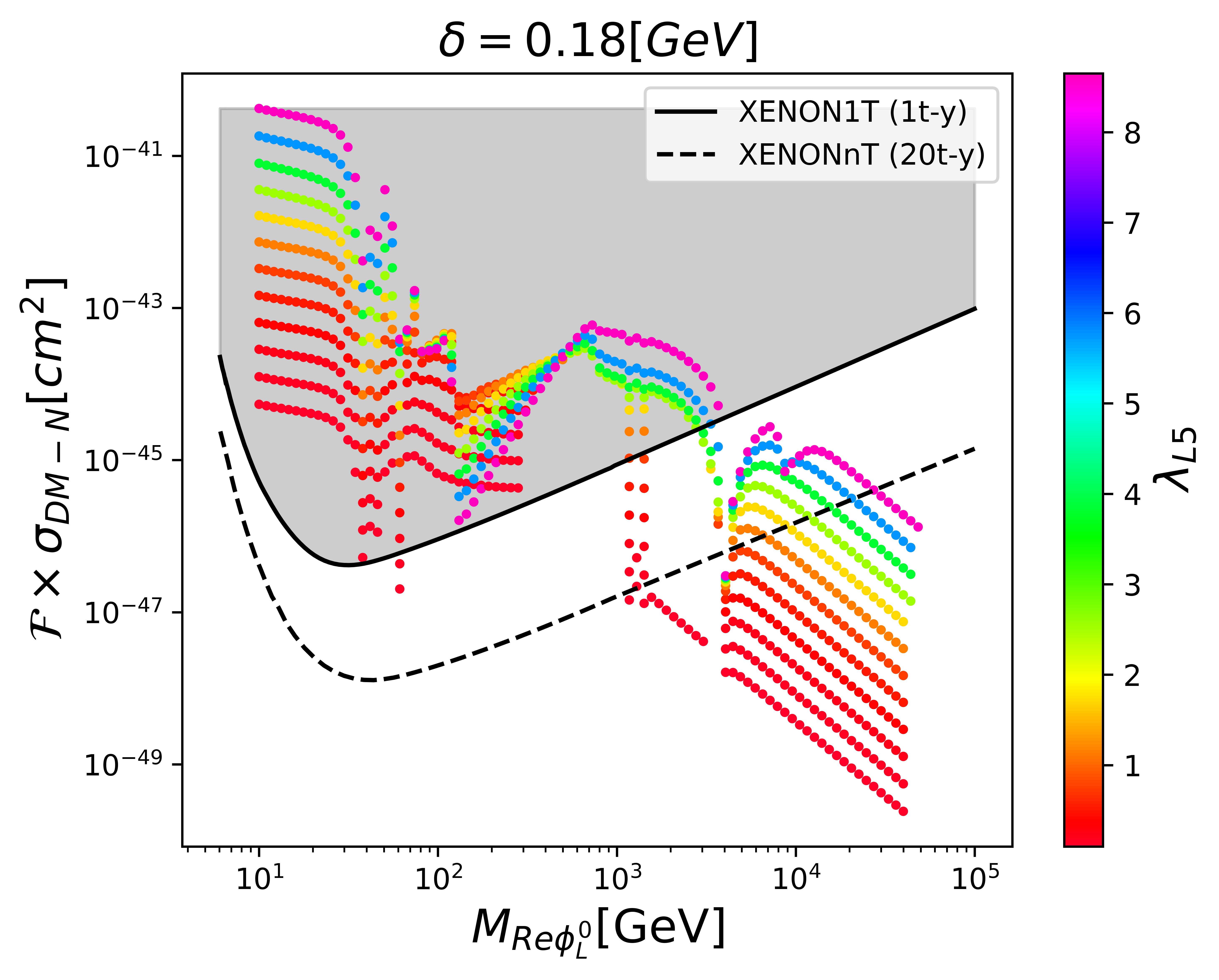}
        \includegraphics[width=0.45\textwidth]{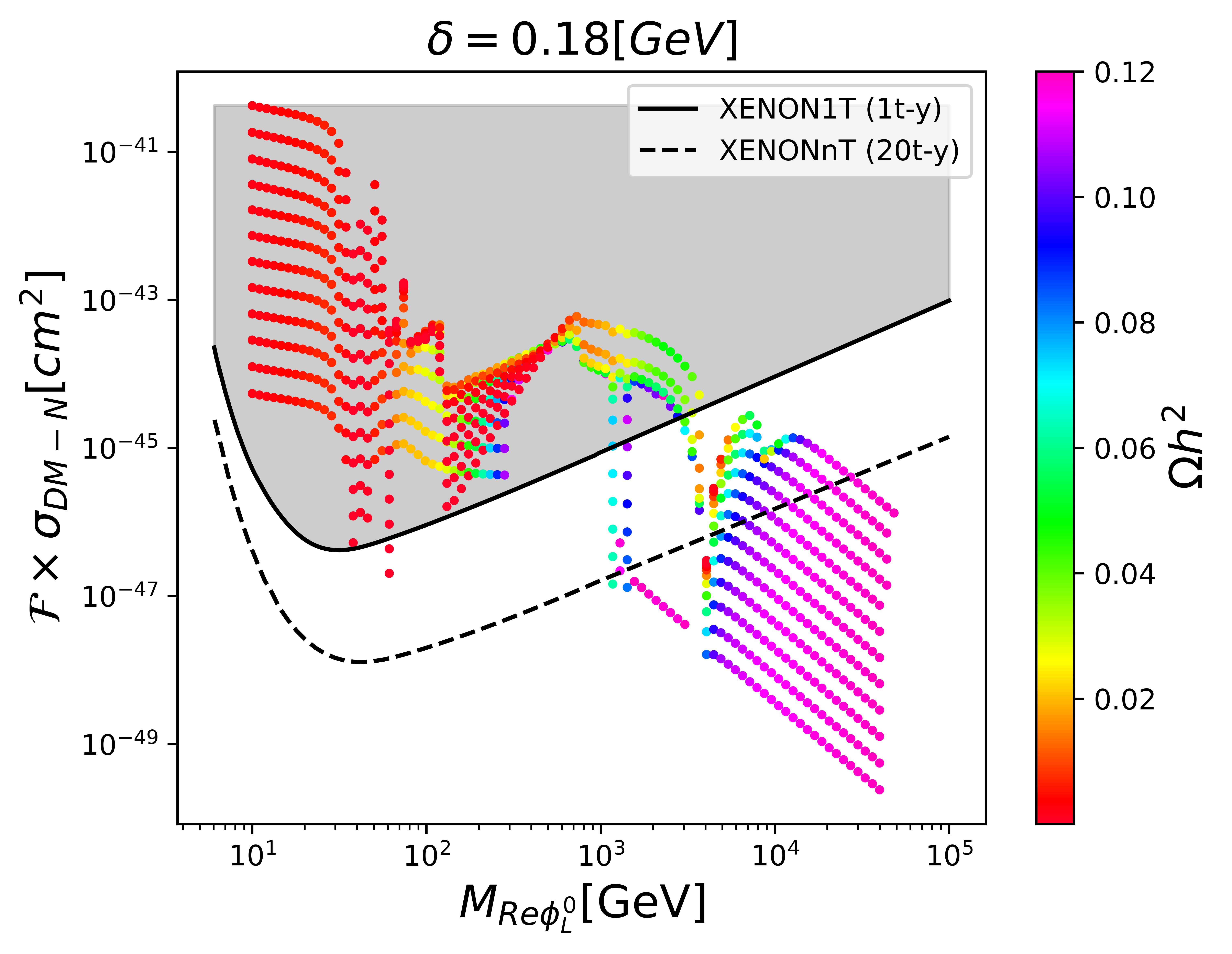}
    \caption{Direct detection prediction for our model. The solid black line corresponds to the exclusion limits reported by XENON1T \cite{XENON1T} while the dashed black line shows the expected sensitivity at XENONnT at 20 ton-years \cite{XENONnT}. The two plots correspond to the same set of points with the color scale in the left panel corresponding to the variation of $\lambda_{L5}$ whereas the color scale in the right panel corresponds to the variation of the DM relic abundance.}
    \label{direct}
\end{figure}

The structure of the model allows for large interactions between the DM candidate and nucleons, making it possible to probe the model at direct detection facilities. In particular, we are focused on the exclusion limits obtained by XENON collaboration~\cite{XENON1T} and the expected sensitivity at XENONnT~\cite{XENONnT}. The predictions for the direct detection cross section in our model are shown in Figure~\ref{direct}. Here, the solid black line corresponds to the exclusion limits reported by XENON1T while the dashed black line shows the expected sensitivity at XENONnT at 20 ton-years. The two plots correspond to the same set of points with the color scale in the left panel corresponding to the variation of $\lambda_{L5}$ whereas the color scale in the right panel corresponds to the variation of the DM relic abundance. As can be seen from Figure~\ref{direct}, the XENON1T bound  excludes masses of the DM below $\sim 100$ GeV. Part of the surviving parameter space lie between the current sensitivity and the expected sensitivity reach.

\subsection{Dark matter production at colliders}

Due to the nature of the present work, we limit ourselves to mentioning some of the interesting possibilities for probing this model at particle collider facilities, in order to define some guidelines for future works related to this topic.

First of all, it is worth mentioning that the low energy realisation of the model resembles an inert doublet model (IDM) with an enriched scalar sector. Therefore, previous studies of the IDM can be a starting point for designing a search strategy\cite{Goudelis:2013uca,Poulose:2016lvz,Belanger:2015kga,Dutta:2017lny,Belyaev:2016lok}. For instance, the production of the neutral scalars along one of the $\mathbb{Z}_2$-even scalars could be probed in mono-Higgs searches \cite{Carpenter:2013xra}. On the other hand, charged scalar double production can be achieved via Drell-Yan and vector boson fusion. The charged production is specially appealing in the light of the favoured mass splitting of $\delta=0.18$ GeV, which allows the decay $\phi_{L}^+\to Re\phi_L^0+ \pi^+$ opening the possibility for probing the model using charged disappearing tracks \cite{Belyaev:2020wok}. However, the model saturates the relic density at a few dozens of TeV making it impossible to produce DM particles at the LHC. The only chance to probe the model as a solution to the DM problem is to study its implications at the FCC. 

\section{Conclusions}
\label{conclusions} 

We have built a renormalizable extended Left-Right symmetric model with an additional global $U(1)_{X}$ symmetry whose spontaneous breaking down to a preserved $\mathbb{Z}_2$ symmetry allows the implementation of the one loop level radiative seesaw that generates the masses of the SM charged fermions lighter than the top quark. In our proposed model, the top quark and exotic fermions obtain tree level masses and the remnant $\mathbb{Z}_2$ symmetry also allows the existence of stable scalar Dark Matter candidates and guarantees two loop level masses for light active neutrinos, arising from a double seesaw mechanism, with the Dirac mass submatrix being generated at one loop level. Our model is capable of successfully explaining and accommodating the observed SM fermion mass hierarchy and the tiny values of the light active neutrino masses. Furthermore, the model under consideration successfully complies with the current constraints from the muon anomalous magnetic moment, the neutrino less double beta decay as well as with those arising from the Dark Matter relic abundance, and its direct and indirect detection.

\section{Acknowledgments}
A.E.C.H., S.K. and I.S. are supported by ANID-Chile FONDECYT 1210378, ANID-Chile
FONDECYT 1180232, ANID-Chile FONDECYT 1230160, ANID-Chile FONDECYT 3150472, ANID PIA/APOYO AFB230003 and
Milenio-ANID-ICN2019\_044. P.A.C. also is supported by Milenio-ANID-ICN2019\_044. V.K.N. is supported by ANID-Chile Fondecyt Postdoctoral grant 3220005. R.P.~is supported in part by the Swedish Research Council grant, contract number 2016-05996, as well as by the European Research Council (ERC) under the European Union's Horizon 2020 research and innovation program (grant agreement No 668679).

\appendix

\bibliographystyle{utphys}
\bibliography{LRdoubleseesaw}

\providecommand{\href}[2]{#2}\begingroup\raggedright\begin{thebibliography}{10}

\bibitem{Pati:1974yy}
J.~C. Pati and A.~Salam, ``{Lepton Number as the Fourth Color},''
  \href{http://dx.doi.org/10.1103/PhysRevD.10.275,
  10.1103/PhysRevD.11.703.2}{{\em Phys. Rev.} {\bfseries D10} (1974) 275--289}.
[Erratum: Phys. Rev.D11,703(1975)].

\bibitem{Mohapatra:1974gc}
R.~N. Mohapatra and J.~C. Pati, ``{A Natural Left-Right Symmetry},''
\href{http://dx.doi.org/10.1103/PhysRevD.11.2558}{{\em Phys. Rev.} {\bfseries
  D11} (1975) 2558}.

\bibitem{Davidson:1987mh}
A.~Davidson and K.~C. Wali, ``{Universal Seesaw Mechanism?},''
\href{http://dx.doi.org/10.1103/PhysRevLett.59.393}{{\em Phys. Rev. Lett.}
  {\bfseries 59} (1987) 393}.

\bibitem{deAlmeida:2010qb}
F.~M.~L. de~Almeida, Y.~A. Coutinho, J.~A. Martins~Simoes, A.~J. Ramalho,
  L.~Ribeiro~Pinto, S.~Wulck, and M.~A.~B. do~Vale, ``{Double seesaw mechanism
  in a left-right symmetric model with TeV neutrinos},''
  \href{http://dx.doi.org/10.1103/PhysRevD.81.053005}{{\em Phys. Rev. D}
  {\bfseries 81} (2010) 053005},
  \href{http://arxiv.org/abs/1001.2162}{{\ttfamily arXiv:1001.2162 [hep-ph]}}.

\bibitem{Gu:2010xc}
P.-H. Gu and U.~Sarkar, ``{Leptogenesis with Linear, Inverse or Double
  Seesaw},'' \href{http://dx.doi.org/10.1016/j.physletb.2010.09.062}{{\em
  Phys.Lett.} {\bfseries B694} (2010) 226--232},
\href{http://arxiv.org/abs/1007.2323}{{\ttfamily arXiv:1007.2323 [hep-ph]}}.

\bibitem{CarcamoHernandez:2018hst}
A.~E. Cárcamo~Hernández, S.~Kovalenko, J.~W.~F. Valle, and C.~A.
  Vaquera-Araujo, ``{Neutrino predictions from a left-right symmetric flavored
  extension of the standard model},''
  \href{http://dx.doi.org/10.1007/JHEP02(2019)065}{{\em JHEP} {\bfseries 02}
  (2019) 065},
\href{http://arxiv.org/abs/1811.03018}{{\ttfamily arXiv:1811.03018 [hep-ph]}}.

\bibitem{Dekens:2014ina}
W.~Dekens and D.~Boer, ``{Viability of minimal left–right models with
  discrete symmetries},''
  \href{http://dx.doi.org/10.1016/j.nuclphysb.2014.10.025}{{\em Nucl. Phys.}
  {\bfseries B889} (2014) 727--756},
\href{http://arxiv.org/abs/1409.4052}{{\ttfamily arXiv:1409.4052 [hep-ph]}}.

\bibitem{Nomura:2016run}
T.~Nomura, H.~Okada, and Y.~Orikasa, ``{Radiative neutrino mass in alternative
  left–right model},''
  \href{http://dx.doi.org/10.1140/epjc/s10052-017-4657-4}{{\em Eur. Phys. J.}
  {\bfseries C77} no.~2, (2017) 103},
\href{http://arxiv.org/abs/1602.08302}{{\ttfamily arXiv:1602.08302 [hep-ph]}}.

\bibitem{Brdar:2018sbk}
V.~Brdar and A.~Y. Smirnov, ``{Low Scale Left-Right Symmetry and Naturally
  Small Neutrino Mass},'' \href{http://dx.doi.org/10.1007/JHEP02(2019)045}{{\em
  JHEP} {\bfseries 02} (2019) 045},
\href{http://arxiv.org/abs/1809.09115}{{\ttfamily arXiv:1809.09115 [hep-ph]}}.

\bibitem{Ma:2020lnm}
E.~Ma, ``{Universal Scotogenic Fermion Masses in Left-Right Gauge Model},''
  \href{http://dx.doi.org/10.1016/j.nuclphysb.2021.115406}{{\em Nucl. Phys.}
  {\bfseries B967} (2021) 115406},
\href{http://arxiv.org/abs/2012.03128}{{\ttfamily arXiv:2012.03128 [hep-ph]}}.

\bibitem{Babu:2020bgz}
K.~S. Babu and A.~Thapa, ``{Left-Right Symmetric Model without Higgs
  Triplets},''
\href{http://arxiv.org/abs/2012.13420}{{\ttfamily arXiv:2012.13420 [hep-ph]}}.

\bibitem{Hernandez:2021uxx}
A.~E.~C. Hern\'andez and I.~Schmidt, ``{A renormalizable left-right symmetric
  model with low scale seesaw mechanisms},''
  \href{http://dx.doi.org/10.1016/j.nuclphysb.2022.115696}{{\em Nucl. Phys. B}
  {\bfseries 976} (2022) 115696},
  \href{http://arxiv.org/abs/2101.02718}{{\ttfamily arXiv:2101.02718
  [hep-ph]}}.

\bibitem{Bonilla:2023wok}
C.~Bonilla, A.~E. Carcamo~Hernandez, S.~Kovalenko, H.~Lee, R.~Pasechnik, and
  I.~Schmidt, ``{Fermion mass hierarchy in an extended left-right symmetric
  model},'' \href{http://dx.doi.org/10.1007/JHEP12(2023)075}{{\em JHEP}
  {\bfseries 12} (2023) 075}, \href{http://arxiv.org/abs/2305.11967}{{\ttfamily
  arXiv:2305.11967 [hep-ph]}}.

\bibitem{Babu:1985gi}
K.~S. Babu, X.-G. He, and S.~Pakvasa, ``{Neutrino Masses and Proton Decay Modes
  in SU(3) X SU(3) X SU(3) Trinification},''
  \href{http://dx.doi.org/10.1103/PhysRevD.33.763}{{\em Phys. Rev. D}
  {\bfseries 33} (1986) 763}.

\bibitem{Choi:2003ag}
K.-S. Choi and J.~E. Kim, ``{Three family Z(3) orbifold trinification, MSSM and
  doublet triplet splitting problem},''
  \href{http://dx.doi.org/10.1016/j.physletb.2003.06.036}{{\em Phys. Lett. B}
  {\bfseries 567} (2003) 87--92},
  \href{http://arxiv.org/abs/hep-ph/0305002}{{\ttfamily arXiv:hep-ph/0305002}}.

\bibitem{Willenbrock:2003ca}
S.~Willenbrock, ``{Triplicated trinification},''
  \href{http://dx.doi.org/10.1016/S0370-2693(03)00419-2}{{\em Phys. Lett. B}
  {\bfseries 561} (2003) 130--134},
  \href{http://arxiv.org/abs/hep-ph/0302168}{{\ttfamily arXiv:hep-ph/0302168}}.

\bibitem{Frampton:2004vw}
P.~H. Frampton and R.~N. Mohapatra, ``{Possible gauge theoretic origin for
  quark-lepton complementarity},''
  \href{http://dx.doi.org/10.1088/1126-6708/2005/01/025}{{\em JHEP} {\bfseries
  01} (2005) 025}, \href{http://arxiv.org/abs/hep-ph/0407139}{{\ttfamily
  arXiv:hep-ph/0407139}}.

\bibitem{Kim:2004pe}
J.~E. Kim, ``{Trinification with sin**2 theta(W) = 3/8 and seesaw neutrino
  mass},'' \href{http://dx.doi.org/10.1016/j.physletb.2004.04.017}{{\em Phys.
  Lett. B} {\bfseries 591} (2004) 119--126},
  \href{http://arxiv.org/abs/hep-ph/0403196}{{\ttfamily arXiv:hep-ph/0403196}}.

\bibitem{Carone:2004rp}
C.~D. Carone and J.~M. Conroy, ``{Higgsless GUT breaking and trinification},''
  \href{http://dx.doi.org/10.1103/PhysRevD.70.075013}{{\em Phys. Rev. D}
  {\bfseries 70} (2004) 075013},
  \href{http://arxiv.org/abs/hep-ph/0407116}{{\ttfamily arXiv:hep-ph/0407116}}.

\bibitem{Sayre:2006ma}
J.~Sayre, S.~Wiesenfeldt, and S.~Willenbrock, ``{Minimal trinification},''
  \href{http://dx.doi.org/10.1103/PhysRevD.73.035013}{{\em Phys. Rev. D}
  {\bfseries 73} (2006) 035013},
  \href{http://arxiv.org/abs/hep-ph/0601040}{{\ttfamily arXiv:hep-ph/0601040}}.

\bibitem{Leontaris:2008mm}
G.~K. Leontaris, ``{A U(3)(C) x U(3)(L) x U(3)(R) gauge symmetry from
  intersecting D-branes},''
  \href{http://dx.doi.org/10.1142/S0217751X0804055X}{{\em Int. J. Mod. Phys. A}
  {\bfseries 23} (2008) 2055--2066},
  \href{http://arxiv.org/abs/0802.4301}{{\ttfamily arXiv:0802.4301 [hep-ph]}}.

\bibitem{Cauet:2010ng}
C.~Cauet, H.~Pas, S.~Wiesenfeldt, H.~Pas, and S.~Wiesenfeldt, ``{Trinification,
  the Hierarchy Problem and Inverse Seesaw Neutrino Masses},''
  \href{http://dx.doi.org/10.1103/PhysRevD.83.093008}{{\em Phys. Rev. D}
  {\bfseries 83} (2011) 093008},
  \href{http://arxiv.org/abs/1012.4083}{{\ttfamily arXiv:1012.4083 [hep-ph]}}.

\bibitem{Stech:2012zr}
B.~Stech, ``{The mass of the Higgs boson in the trinification subgroup of
  E6},'' \href{http://dx.doi.org/10.1103/PhysRevD.86.055003}{{\em Phys. Rev. D}
  {\bfseries 86} (2012) 055003},
  \href{http://arxiv.org/abs/1206.4233}{{\ttfamily arXiv:1206.4233 [hep-ph]}}.

\bibitem{Stech:2014tla}
B.~Stech, ``{Trinification Phenomenology and the structure of Higgs Bosons},''
  \href{http://dx.doi.org/10.1007/JHEP08(2014)139}{{\em JHEP} {\bfseries 08}
  (2014) 139}, \href{http://arxiv.org/abs/1403.2714}{{\ttfamily arXiv:1403.2714
  [hep-ph]}}.

\bibitem{Hetzel:2015bla}
J.~Hetzel and B.~Stech, ``{Low-energy phenomenology of trinification: an
  effective left-right-symmetric model},''
  \href{http://dx.doi.org/10.1103/PhysRevD.91.055026}{{\em Phys. Rev. D}
  {\bfseries 91} (2015) 055026},
  \href{http://arxiv.org/abs/1502.00919}{{\ttfamily arXiv:1502.00919
  [hep-ph]}}.

\bibitem{Hetzel:2015cca}
J.~Hetzel, \href{http://dx.doi.org/10.11588/heidok.00018259}{{\em
  {Phenomenology of a left-right-symmetric model inspired by the trinification
  model}}}.
\newblock PhD thesis, U. Heidelberg (main), 2015.
\newblock \href{http://arxiv.org/abs/1504.06739}{{\ttfamily arXiv:1504.06739
  [hep-ph]}}.

\bibitem{Pelaggi:2015kna}
G.~M. Pelaggi, A.~Strumia, and S.~Vignali, ``{Totally asymptotically free
  trinification},'' \href{http://dx.doi.org/10.1007/JHEP08(2015)130}{{\em JHEP}
  {\bfseries 08} (2015) 130}, \href{http://arxiv.org/abs/1507.06848}{{\ttfamily
  arXiv:1507.06848 [hep-ph]}}.

\bibitem{Camargo-Molina:2016yqm}
J.~E. Camargo-Molina, A.~P. Morais, A.~Ordell, R.~Pasechnik, M.~O.~P. Sampaio,
  and J.~Wess\'en, ``{Reviving trinification models through an E6 -extended
  supersymmetric GUT},''
  \href{http://dx.doi.org/10.1103/PhysRevD.95.075031}{{\em Phys. Rev. D}
  {\bfseries 95} no.~7, (2017) 075031},
  \href{http://arxiv.org/abs/1610.03642}{{\ttfamily arXiv:1610.03642
  [hep-ph]}}.

\bibitem{Reig:2016vtf}
M.~Reig, J.~W.~F. Valle, and C.~A. Vaquera-Araujo, ``{Three-family left-right
  symmetry with low-scale seesaw mechanism},''
  \href{http://dx.doi.org/10.1007/JHEP05(2017)100}{{\em JHEP} {\bfseries 05}
  (2017) 100}, \href{http://arxiv.org/abs/1611.04571}{{\ttfamily
  arXiv:1611.04571 [hep-ph]}}.

\bibitem{Camargo-Molina:2016bwm}
J.~E. Camargo-Molina, A.~P. Morais, R.~Pasechnik, and J.~Wess\'en, ``{On a
  radiative origin of the Standard Model from Trinification},''
  \href{http://dx.doi.org/10.1007/JHEP09(2016)129}{{\em JHEP} {\bfseries 09}
  (2016) 129}, \href{http://arxiv.org/abs/1606.03492}{{\ttfamily
  arXiv:1606.03492 [hep-ph]}}.

\bibitem{Reig:2016tuk}
M.~Reig, J.~W.~F. Valle, and C.~A. Vaquera-Araujo, ``{Unifying
  left\textendash{}right symmetry and 331 electroweak theories},''
  \href{http://dx.doi.org/10.1016/j.physletb.2016.12.049}{{\em Phys. Lett. B}
  {\bfseries 766} (2017) 35--40},
  \href{http://arxiv.org/abs/1611.02066}{{\ttfamily arXiv:1611.02066
  [hep-ph]}}.

\bibitem{Hati:2017aez}
C.~Hati, S.~Patra, M.~Reig, J.~W.~F. Valle, and C.~A. Vaquera-Araujo,
  ``{Towards gauge coupling unification in left-right symmetric
  $\mathrm{SU(3)_c \times SU(3)_L \times SU(3)_R \times U(1)_{X}}$ theories},''
  \href{http://dx.doi.org/10.1103/PhysRevD.96.015004}{{\em Phys. Rev. D}
  {\bfseries 96} no.~1, (2017) 015004},
  \href{http://arxiv.org/abs/1703.09647}{{\ttfamily arXiv:1703.09647
  [hep-ph]}}.

\bibitem{Camargo-Molina:2017kxd}
J.~E. Camargo-Molina, A.~P. Morais, A.~Ordell, R.~Pasechnik, and J.~Wess\'en,
  ``{Scale hierarchies, symmetry breaking and particle spectra in SU(3)-family
  extended SUSY trinification},''
  \href{http://dx.doi.org/10.1103/PhysRevD.99.035041}{{\em Phys. Rev. D}
  {\bfseries 99} no.~3, (2019) 035041},
  \href{http://arxiv.org/abs/1711.05199}{{\ttfamily arXiv:1711.05199
  [hep-ph]}}.

\bibitem{Dong:2017zxo}
P.~V. Dong, D.~T. Huong, F.~S. Queiroz, J.~W.~F. Valle, and C.~A.
  Vaquera-Araujo, ``{The Dark Side of Flipped Trinification},''
  \href{http://dx.doi.org/10.1007/JHEP04(2018)143}{{\em JHEP} {\bfseries 04}
  (2018) 143}, \href{http://arxiv.org/abs/1710.06951}{{\ttfamily
  arXiv:1710.06951 [hep-ph]}}.

\bibitem{Wang:2018yer}
Z.-W. Wang, A.~Al~Balushi, R.~Mann, and H.-M. Jiang, ``{Safe Trinification},''
  \href{http://dx.doi.org/10.1103/PhysRevD.99.115017}{{\em Phys. Rev. D}
  {\bfseries 99} no.~11, (2019) 115017},
  \href{http://arxiv.org/abs/1812.11085}{{\ttfamily arXiv:1812.11085
  [hep-ph]}}.

\bibitem{Dinh:2019jdg}
D.~N. Dinh, D.~T. Huong, N.~T. Duy, N.~T. Nhuan, L.~D. Thien, and P.~Van~Dong,
  ``{Flavor changing in the flipped trinification},''
  \href{http://dx.doi.org/10.1103/PhysRevD.99.055005}{{\em Phys. Rev. D}
  {\bfseries 99} no.~5, (2019) 055005},
  \href{http://arxiv.org/abs/1901.07969}{{\ttfamily arXiv:1901.07969
  [hep-ph]}}.

\bibitem{Aranda:2020fkj}
A.~Aranda, F.~J. de~Anda, A.~P. Morais, and R.~Pasechnik, ``{Gauge Couplings
  Evolution from the Standard Model, through Pati\textendash{}Salam Theory,
  into E$_{8}$ Unification of Families and Forces},''
  \href{http://dx.doi.org/10.3390/universe9020090}{{\em Universe} {\bfseries 9}
  no.~2, (2023) 90}, \href{http://arxiv.org/abs/2011.13902}{{\ttfamily
  arXiv:2011.13902 [hep-ph]}}.

\bibitem{Morais:2020ypd}
A.~P. Morais, R.~Pasechnik, and W.~Porod, ``{Prospects for new physics from
  gauge left-right-colour-family grand unification hypothesis},''
  \href{http://dx.doi.org/10.1140/epjc/s10052-020-08710-4}{{\em Eur. Phys. J.
  C} {\bfseries 80} no.~12, (2020) 1162},
  \href{http://arxiv.org/abs/2001.06383}{{\ttfamily arXiv:2001.06383
  [hep-ph]}}.

\bibitem{Morais:2020odg}
A.~P. Morais, R.~Pasechnik, and W.~Porod, ``{Grand Unified Origin of Gauge
  Interactions and Families Replication in the Standard Model},''
  \href{http://dx.doi.org/10.3390/universe7120461}{{\em Universe} {\bfseries 7}
  no.~12, (2021) 461}, \href{http://arxiv.org/abs/2001.04804}{{\ttfamily
  arXiv:2001.04804 [hep-ph]}}.

\bibitem{CarcamoHernandez:2020owa}
A.~E. C\'arcamo~Hern\'andez, D.~T. Huong, S.~Kovalenko, A.~P. Morais,
  R.~Pasechnik, and I.~Schmidt, ``{How low-scale trinification sheds light in
  the flavor hierarchies, neutrino puzzle, dark matter, and leptogenesis},''
  \href{http://dx.doi.org/10.1103/PhysRevD.102.095003}{{\em Phys. Rev. D}
  {\bfseries 102} no.~9, (2020) 095003},
  \href{http://arxiv.org/abs/2004.11450}{{\ttfamily arXiv:2004.11450
  [hep-ph]}}.

\bibitem{Aranda:2021bvg}
A.~Aranda, F.~J. de~Anda, A.~P. Morais, and R.~Pasechnik, ``{Can $E_8$
  unification at low energies be consistent with proton decay?},''
  \href{http://arxiv.org/abs/2107.05421}{{\ttfamily arXiv:2107.05421
  [hep-ph]}}.

\bibitem{Aranda:2021eyn}
A.~Aranda, F.~J. de~Anda, A.~P. Morais, and R.~Pasechnik, ``{Sculpting the
  Standard Model from low-scale gauge-Higgs-matter E8 grand unification in ten
  dimensions},'' \href{http://dx.doi.org/10.1016/j.nuclphysb.2023.116266}{{\em
  Nucl. Phys. B} {\bfseries 993} (2023) 116266},
  \href{http://arxiv.org/abs/2107.05495}{{\ttfamily arXiv:2107.05495
  [hep-ph]}}.

\bibitem{Patra:2023ltl}
S.~Patra, S.~T. Petcov, P.~Pritimita, and P.~Sahu, ``{Neutrinoless double beta
  decay in a left-right symmetric model with a double seesaw mechanism},''
  \href{http://dx.doi.org/10.1103/PhysRevD.107.075037}{{\em Phys. Rev. D}
  {\bfseries 107} no.~7, (2023) 075037},
  \href{http://arxiv.org/abs/2302.14538}{{\ttfamily arXiv:2302.14538
  [hep-ph]}}.

\bibitem{Patel:2023voj}
U.~Patel, P.~Adarsh, S.~Patra, and P.~Sahu, ``{Leptogenesis in a Left-Right
  Symmetric Model with double seesaw},''
  \href{http://dx.doi.org/10.1007/JHEP03(2024)029}{{\em JHEP} {\bfseries 03}
  (2024) 029}, \href{http://arxiv.org/abs/2310.09337}{{\ttfamily
  arXiv:2310.09337 [hep-ph]}}.

\bibitem{Xing:2020ijf}
Z.-z. Xing, ``{Flavor structures of charged fermions and massive neutrinos},''
  \href{http://dx.doi.org/10.1016/j.physrep.2020.02.001}{{\em Phys. Rept.}
  {\bfseries 854} (2020) 1--147},
  \href{http://arxiv.org/abs/1909.09610}{{\ttfamily arXiv:1909.09610
  [hep-ph]}}.

\bibitem{ParticleDataGroup:2022pth}
{\bfseries Particle Data Group} Collaboration, R.~L. Workman {\em et~al.},
  ``{Review of Particle Physics},''
  \href{http://dx.doi.org/10.1093/ptep/ptac097}{{\em PTEP} {\bfseries 2022}
  (2022) 083C01}.

\bibitem{Barr:2003nn}
S.~M. Barr, ``{A Different seesaw formula for neutrino masses},''
  \href{http://dx.doi.org/10.1103/PhysRevLett.92.101601}{{\em Phys. Rev. Lett.}
  {\bfseries 92} (2004) 101601},
  \href{http://arxiv.org/abs/hep-ph/0309152}{{\ttfamily arXiv:hep-ph/0309152}}.

\bibitem{Casas:2001sr}
J.~Casas and A.~Ibarra, ``{Oscillating neutrinos and muon ---> e, gamma},''
  \href{http://dx.doi.org/10.1016/S0550-3213(01)00475-8}{{\em Nucl.Phys.}
  {\bfseries B618} (2001) 171--204},
\href{http://arxiv.org/abs/hep-ph/0103065}{{\ttfamily arXiv:hep-ph/0103065
  [hep-ph]}}.

\bibitem{Ibarra:2003up}
A.~Ibarra and G.~G. Ross, ``{Neutrino phenomenology: The Case of two
  right-handed neutrinos},''
  \href{http://dx.doi.org/10.1016/j.physletb.2004.04.037}{{\em Phys.Lett.}
  {\bfseries B591} (2004) 285--296},
\href{http://arxiv.org/abs/hep-ph/0312138}{{\ttfamily arXiv:hep-ph/0312138
  [hep-ph]}}.

\bibitem{Diaz:2002uk}
R.~A. Diaz, R.~Martinez, and J.~A. Rodriguez, ``{Phenomenology of lepton flavor
  violation in 2HDM(3) from (g-2)(mu) and leptonic decays},''
  \href{http://dx.doi.org/10.1103/PhysRevD.67.075011}{{\em Phys. Rev.}
  {\bfseries D67} (2003) 075011},
\href{http://arxiv.org/abs/hep-ph/0208117}{{\ttfamily arXiv:hep-ph/0208117
  [hep-ph]}}.

\bibitem{Jegerlehner:2009ry}
F.~Jegerlehner and A.~Nyffeler, ``{The Muon g-2},''
  \href{http://dx.doi.org/10.1016/j.physrep.2009.04.003}{{\em Phys. Rept.}
  {\bfseries 477} (2009) 1--110},
\href{http://arxiv.org/abs/0902.3360}{{\ttfamily arXiv:0902.3360 [hep-ph]}}.

\bibitem{Kelso:2014qka}
C.~Kelso, H.~N. Long, R.~Martinez, and F.~S. Queiroz, ``{Connection of
  $g-2_{\mu}$, electroweak, dark matter, and collider constraints on 331
  models},'' \href{http://dx.doi.org/10.1103/PhysRevD.90.113011}{{\em Phys.
  Rev.} {\bfseries D90} no.~11, (2014) 113011},
\href{http://arxiv.org/abs/1408.6203}{{\ttfamily arXiv:1408.6203 [hep-ph]}}.

\bibitem{Lindner:2016bgg}
M.~Lindner, M.~Platscher, and F.~S. Queiroz, ``{A Call for New Physics : The
  Muon Anomalous Magnetic Moment and Lepton Flavor Violation},''
  \href{http://dx.doi.org/10.1016/j.physrep.2017.12.001}{{\em Phys. Rept.}
  {\bfseries 731} (2018) 1--82},
\href{http://arxiv.org/abs/1610.06587}{{\ttfamily arXiv:1610.06587 [hep-ph]}}.

\bibitem{Kowalska:2017iqv}
K.~Kowalska and E.~M. Sessolo, ``{Expectations for the muon g-2 in simplified
  models with dark matter},''
  \href{http://dx.doi.org/10.1007/JHEP09(2017)112}{{\em JHEP} {\bfseries 09}
  (2017) 112},
\href{http://arxiv.org/abs/1707.00753}{{\ttfamily arXiv:1707.00753 [hep-ph]}}.

\bibitem{Muong-2:2023cdq}
{\bfseries Muon g-2} Collaboration, D.~P. Aguillard {\em et~al.},
  ``{Measurement of the Positive Muon Anomalous Magnetic Moment to 0.20~ppm},''
  \href{http://dx.doi.org/10.1103/PhysRevLett.131.161802}{{\em Phys. Rev.
  Lett.} {\bfseries 131} no.~16, (2023) 161802},
  \href{http://arxiv.org/abs/2308.06230}{{\ttfamily arXiv:2308.06230
  [hep-ex]}}.

\bibitem{Hirsch:1996qw}
M.~Hirsch, H.~Klapdor-Kleingrothaus, and O.~Panella, ``{Double beta decay in
  left-right symmetric models},''
  \href{http://dx.doi.org/10.1016/0370-2693(96)00185-2}{{\em Phys. Lett. B}
  {\bfseries 374} (1996) 7--12},
  \href{http://arxiv.org/abs/hep-ph/9602306}{{\ttfamily arXiv:hep-ph/9602306}}.

\bibitem{Tello:2010am}
V.~Tello, M.~Nemevsek, F.~Nesti, G.~Senjanovic, and F.~Vissani, ``{Left-Right
  Symmetry: from LHC to Neutrinoless Double Beta Decay},''
  \href{http://dx.doi.org/10.1103/PhysRevLett.106.151801}{{\em Phys.Rev.Lett.}
  {\bfseries 106} (2011) 151801},
\href{http://arxiv.org/abs/1011.3522}{{\ttfamily arXiv:1011.3522 [hep-ph]}}.

\bibitem{Chakrabortty:2012mh}
J.~Chakrabortty, H.~Z. Devi, S.~Goswami, and S.~Patra, ``{Neutrinoless
  double-$\beta$ decay in TeV scale Left-Right symmetric models},''
  \href{http://dx.doi.org/10.1007/JHEP08(2012)008}{{\em JHEP} {\bfseries 1208}
  (2012) 008},
\href{http://arxiv.org/abs/1204.2527}{{\ttfamily arXiv:1204.2527 [hep-ph]}}.

\bibitem{Barry:2013xxa}
J.~Barry and W.~Rodejohann, ``{Lepton number and flavour violation in TeV-scale
  left-right symmetric theories with large left-right mixing},''
  \href{http://dx.doi.org/10.1007/JHEP09(2013)153}{{\em JHEP} {\bfseries 09}
  (2013) 153},
\href{http://arxiv.org/abs/1303.6324}{{\ttfamily arXiv:1303.6324 [hep-ph]}}.

\bibitem{Dev:2014xea}
P.~S. Bhupal~Dev, S.~Goswami, and M.~Mitra, ``{TeV Scale Left-Right Symmetry
  and Large Mixing Effects in Neutrinoless Double Beta Decay},''
  \href{http://dx.doi.org/10.1103/PhysRevD.91.113004}{{\em Phys. Rev.}
  {\bfseries D91} no.~11, (2015) 113004},
\href{http://arxiv.org/abs/1405.1399}{{\ttfamily arXiv:1405.1399 [hep-ph]}}.

\bibitem{Awasthi:2015ota}
R.~L. Awasthi, P.~S.~B. Dev, and M.~Mitra, ``{Implications of the Diboson
  Excess for Neutrinoless Double Beta Decay and Lepton Flavor Violation in TeV
  Scale Left Right Symmetric Model},''
  \href{http://dx.doi.org/10.1103/PhysRevD.93.011701}{{\em Phys. Rev.}
  {\bfseries D93} no.~1, (2016) 011701},
\href{http://arxiv.org/abs/1509.05387}{{\ttfamily arXiv:1509.05387 [hep-ph]}}.

\bibitem{Bambhaniya:2015ipg}
G.~Bambhaniya, P.~S.~B. Dev, S.~Goswami, and M.~Mitra, ``{The Scalar Triplet
  Contribution to Lepton Flavour Violation and Neutrinoless Double Beta Decay
  in Left-Right Symmetric Model},''
  \href{http://dx.doi.org/10.1007/JHEP04(2016)046}{{\em JHEP} {\bfseries 04}
  (2016) 046},
\href{http://arxiv.org/abs/1512.00440}{{\ttfamily arXiv:1512.00440 [hep-ph]}}.

\bibitem{Bonilla:2016fqd}
C.~Bonilla, M.~E. Krauss, T.~Opferkuch, and W.~Porod, ``{Perspectives for
  Detecting Lepton Flavour Violation in Left-Right Symmetric Models},''
  \href{http://dx.doi.org/10.1007/JHEP03(2017)027}{{\em JHEP} {\bfseries 03}
  (2017) 027}, \href{http://arxiv.org/abs/1611.07025}{{\ttfamily
  arXiv:1611.07025 [hep-ph]}}.

\bibitem{Awasthi:2016kbk}
R.~L. Awasthi, A.~Dasgupta, and M.~Mitra, ``{Limiting the effective mass and
  new physics parameters from $0\nu\beta\beta$},''
  \href{http://dx.doi.org/10.1103/PhysRevD.94.073003}{{\em Phys. Rev. D}
  {\bfseries 94} no.~7, (2016) 073003},
  \href{http://arxiv.org/abs/1607.03835}{{\ttfamily arXiv:1607.03835
  [hep-ph]}}.

\bibitem{Deppisch:2017vne}
F.~F. Deppisch, C.~Hati, S.~Patra, P.~Pritimita, and U.~Sarkar, ``{Neutrinoless
  double beta decay in left-right symmetric models with a universal seesaw
  mechanism},'' \href{http://dx.doi.org/10.1103/PhysRevD.97.035005}{{\em Phys.
  Rev. D} {\bfseries 97} no.~3, (2018) 035005},
  \href{http://arxiv.org/abs/1701.02107}{{\ttfamily arXiv:1701.02107
  [hep-ph]}}.

\bibitem{Borah:2017ldt}
D.~Borah, A.~Dasgupta, and S.~Patra, ``{Neutrinoless double beta decay in
  minimal left--right symmetric model with universal seesaw},''
  \href{http://dx.doi.org/10.1142/S0217751X18501981}{{\em Int. J. Mod. Phys. A}
  {\bfseries 33} no.~35, (2018) 1850198},
  \href{http://arxiv.org/abs/1706.02456}{{\ttfamily arXiv:1706.02456
  [hep-ph]}}.

\bibitem{Borgohain:2017akh}
H.~Borgohain and M.~K. Das, ``{Lepton number violation, lepton flavor
  violation, and baryogenesis in left-right symmetric model},''
  \href{http://dx.doi.org/10.1103/PhysRevD.96.075021}{{\em Phys. Rev. D}
  {\bfseries 96} no.~7, (2017) 075021},
  \href{http://arxiv.org/abs/1709.09542}{{\ttfamily arXiv:1709.09542
  [hep-ph]}}.

\bibitem{Goswami:2020loc}
S.~Goswami and K.~N. Vishnudath, ``{Low energy constraints from absolute
  neutrino mass observables and lepton flavor violation in left-right symmetric
  model},'' \href{http://dx.doi.org/10.1103/PhysRevD.103.055016}{{\em Phys.
  Rev. D} {\bfseries 103} no.~5, (2021) 055016},
  \href{http://arxiv.org/abs/2011.06314}{{\ttfamily arXiv:2011.06314
  [hep-ph]}}.

\bibitem{Agostini:2018tnm}
{\bfseries GERDA} Collaboration, M.~Agostini {\em et~al.}, ``{Improved Limit on
  Neutrinoless Double-$\beta$ Decay of $^{76}$Ge from GERDA Phase II},''
  \href{http://dx.doi.org/10.1103/PhysRevLett.120.132503}{{\em Phys. Rev.
  Lett.} {\bfseries 120} no.~13, (2018) 132503},
\href{http://arxiv.org/abs/1803.11100}{{\ttfamily arXiv:1803.11100 [nucl-ex]}}.

\bibitem{Kharusi:2018eqi}
{\bfseries nEXO} Collaboration, S.~A. Kharusi {\em et~al.}, ``{nEXO
  Pre-Conceptual Design Report},''
\href{http://arxiv.org/abs/1805.11142}{{\ttfamily arXiv:1805.11142
  [physics.ins-det]}}.

\bibitem{Ashry:2013loa}
M.~Ashry and S.~Khalil, ``{Phenomenological aspects of a TeV-scale alternative
  left-right model},'' \href{http://dx.doi.org/10.1103/PhysRevD.91.015009}{{\em
  Phys. Rev. D} {\bfseries 91} no.~1, (2015) 015009},
  \href{http://arxiv.org/abs/1310.3315}{{\ttfamily arXiv:1310.3315 [hep-ph]}}.
  [Addendum: Phys.Rev.D 96, 059901 (2017)].

\bibitem{fr1}
N.~D. Christensen and C.~Duhr, ``{FeynRules} {\textendash} feynman rules made
  easy,'' \href{http://dx.doi.org/10.1016/j.cpc.2009.02.018}{{\em Computer
  Physics Communications} {\bfseries 180} no.~9, (Sep, 2009) 1614--1641}.
  \url{https://doi.org/10.1016%2Fj.cpc.2009.02.018}.

\bibitem{fr2}
A.~Alloul, N.~D. Christensen, C.~Degrande, C.~Duhr, and B.~Fuks, ``{FeynRules}
  ~2.0~{\textemdash} a complete toolbox for tree-level phenomenology,''
  \href{http://dx.doi.org/10.1016/j.cpc.2014.04.012}{{\em Computer Physics
  Communications} {\bfseries 185} no.~8, (Aug, 2014) 2250--2300}.
  \url{https://doi.org/10.1016%2Fj.cpc.2014.04.012}.

\bibitem{micromegas1}
G.~Belanger, F.~Boudjema, A.~Pukhov, and A.~Semenov, ``{micrOMEGAs$\_$3: A
  program for calculating dark matter observables},''
  \href{http://dx.doi.org/10.1016/j.cpc.2013.10.016}{{\em Comput. Phys.
  Commun.} {\bfseries 185} (2014) 960--985},
  \href{http://arxiv.org/abs/1305.0237}{{\ttfamily arXiv:1305.0237 [hep-ph]}}.

\bibitem{micromegas2}
G.~Belanger, F.~Boudjema, A.~Pukhov, and A.~Semenov, ``{Dark matter direct
  detection rate in a generic model with micrOMEGAs 2.2},''
  \href{http://dx.doi.org/10.1016/j.cpc.2008.11.019}{{\em Comput. Phys.
  Commun.} {\bfseries 180} (2009) 747--767},
  \href{http://arxiv.org/abs/0803.2360}{{\ttfamily arXiv:0803.2360 [hep-ph]}}.

\bibitem{micromegas3}
G.~Belanger, F.~Boudjema, A.~Pukhov, and A.~Semenov, ``{MicrOMEGAs 2.0: A
  Program to calculate the relic density of dark matter in a generic model},''
  \href{http://dx.doi.org/10.1016/j.cpc.2006.11.008}{{\em Comput. Phys.
  Commun.} {\bfseries 176} (2007) 367--382},
  \href{http://arxiv.org/abs/hep-ph/0607059}{{\ttfamily arXiv:hep-ph/0607059}}.

\bibitem{ctalims}
{\bfseries CTA} Collaboration, A.~Acharyya {\em et~al.}, ``{Sensitivity of the
  Cherenkov Telescope Array to a dark matter signal from the Galactic
  centre},'' \href{http://dx.doi.org/10.1088/1475-7516/2021/01/057}{{\em JCAP}
  {\bfseries 01} (2021) 057}, \href{http://arxiv.org/abs/2007.16129}{{\ttfamily
  arXiv:2007.16129 [astro-ph.HE]}}.

\bibitem{XENON1T}
{\bfseries XENON} Collaboration, E.~Aprile {\em et~al.}, ``{Dark Matter Search
  Results from a One Ton-Year Exposure of XENON1T},''
  \href{http://dx.doi.org/10.1103/PhysRevLett.121.111302}{{\em Phys. Rev.
  Lett.} {\bfseries 121} no.~11, (2018) 111302},
  \href{http://arxiv.org/abs/1805.12562}{{\ttfamily arXiv:1805.12562
  [astro-ph.CO]}}.

\bibitem{XENONnT}
{\bfseries XENON} Collaboration, E.~Aprile {\em et~al.}, ``{Projected WIMP
  sensitivity of the XENONnT dark matter experiment},''
  \href{http://dx.doi.org/10.1088/1475-7516/2020/11/031}{{\em JCAP} {\bfseries
  11} (2020) 031}, \href{http://arxiv.org/abs/2007.08796}{{\ttfamily
  arXiv:2007.08796 [physics.ins-det]}}.

\bibitem{Goudelis:2013uca}
A.~Goudelis, B.~Herrmann, and O.~St\r{a}l, ``{Dark matter in the Inert Doublet
  Model after the discovery of a Higgs-like boson at the LHC},''
  \href{http://dx.doi.org/10.1007/JHEP09(2013)106}{{\em JHEP} {\bfseries 09}
  (2013) 106}, \href{http://arxiv.org/abs/1303.3010}{{\ttfamily arXiv:1303.3010
  [hep-ph]}}.

\bibitem{Poulose:2016lvz}
P.~Poulose, S.~Sahoo, and K.~Sridhar, ``{Exploring the Inert Doublet Model
  through the dijet plus missing transverse energy channel at the LHC},''
  \href{http://dx.doi.org/10.1016/j.physletb.2016.12.022}{{\em Phys. Lett. B}
  {\bfseries 765} (2017) 300--306},
  \href{http://arxiv.org/abs/1604.03045}{{\ttfamily arXiv:1604.03045
  [hep-ph]}}.

\bibitem{Belanger:2015kga}
G.~Belanger, B.~Dumont, A.~Goudelis, B.~Herrmann, S.~Kraml, and D.~Sengupta,
  ``{Dilepton constraints in the Inert Doublet Model from Run 1 of the LHC},''
  \href{http://dx.doi.org/10.1103/PhysRevD.91.115011}{{\em Phys. Rev. D}
  {\bfseries 91} no.~11, (2015) 115011},
  \href{http://arxiv.org/abs/1503.07367}{{\ttfamily arXiv:1503.07367
  [hep-ph]}}.

\bibitem{Dutta:2017lny}
B.~Dutta, G.~Palacio, J.~D. Ruiz-Alvarez, and D.~Restrepo, ``{Vector Boson
  Fusion in the Inert Doublet Model},''
  \href{http://dx.doi.org/10.1103/PhysRevD.97.055045}{{\em Phys. Rev. D}
  {\bfseries 97} no.~5, (2018) 055045},
  \href{http://arxiv.org/abs/1709.09796}{{\ttfamily arXiv:1709.09796
  [hep-ph]}}.

\bibitem{Belyaev:2016lok}
A.~Belyaev, G.~Cacciapaglia, I.~P. Ivanov, F.~Rojas-Abatte, and M.~Thomas,
  ``{Anatomy of the Inert Two Higgs Doublet Model in the light of the LHC and
  non-LHC Dark Matter Searches},''
  \href{http://dx.doi.org/10.1103/PhysRevD.97.035011}{{\em Phys. Rev. D}
  {\bfseries 97} no.~3, (2018) 035011},
  \href{http://arxiv.org/abs/1612.00511}{{\ttfamily arXiv:1612.00511
  [hep-ph]}}.

\bibitem{Carpenter:2013xra}
L.~Carpenter, A.~DiFranzo, M.~Mulhearn, C.~Shimmin, S.~Tulin, and D.~Whiteson,
  ``{Mono-Higgs-boson: A new collider probe of dark matter},''
  \href{http://dx.doi.org/10.1103/PhysRevD.89.075017}{{\em Phys. Rev. D}
  {\bfseries 89} no.~7, (2014) 075017},
  \href{http://arxiv.org/abs/1312.2592}{{\ttfamily arXiv:1312.2592 [hep-ph]}}.

\bibitem{Belyaev:2020wok}
A.~Belyaev, S.~Prestel, F.~Rojas-Abbate, and J.~Zurita, ``{Probing dark matter
  with disappearing tracks at the LHC},''
  \href{http://dx.doi.org/10.1103/PhysRevD.103.095006}{{\em Phys. Rev. D}
  {\bfseries 103} no.~9, (2021) 095006},
  \href{http://arxiv.org/abs/2008.08581}{{\ttfamily arXiv:2008.08581
  [hep-ph]}}.

\end{thebibliography}\endgroup

\end{document}